\documentclass[twocolumn]{aastex63}

\received{\today}
\revised{\today}
\accepted{\today}
\submitjournal{ApJ}

\shorttitle{\textmyfont{Fe II} and \textmyfont{Ca II} emission in AGNs: Paper I}
\shortauthors{Panda et al.}
\graphicspath{{./}{figures/}}

\newenvironment{myfont}{\fontfamily{lmtt}\selectfont}{\par}
\DeclareTextFontCommand{\textmyfont}{\myfont}
\usepackage{amsmath}
\usepackage{natbib}
\usepackage[caption=false]{subfig}
\usepackage[normalem]{ulem}
\usepackage[T1]{fontenc}
\usepackage{longtable}
\usepackage{soul}
\usepackage{dirtytalk}

\def\kms{\,km\,s$^{-1}$}
\def\ergs{erg\,s$^{-1}$}
\def\hb{{\sc{H}}$\beta$\/}

\def\feii{\rm{Fe \sc{ii}}}
\def\caii{\rm{Ca {\sc{ii}}}}
\def\rfe{R$_{\rm{FeII}}$}
\def\rcat{R$_{\rm{CaT}}$}
\def\cat{\rm{CaT}}
\def\lbol{$L\mathrm{_{bol}}$}
\def\mbh{$M\mathrm{_{BH}}$}

\def\LLEdd{$L\mathrm{_{bol}}/L\mathrm{_{Edd}}$}

\def\zsun{Z$_{\odot}$}

\def\un{$\log \rm{U} - \log \rm{n_{H}}$}
\def\rblr{$R\mathrm{_{BLR}}$}
\def\oi{O {\sc i} $\lambda8446$}
\def\hb{H$\beta$}

\def\n{$\rm{n_{H}}$}
\def\u{$\rm{U}$}

\begin{document}

\title{The CaFe Project: Optical \textmyfont{Fe II} and Near-Infrared \textmyfont{Ca II} triplet emission in active galaxies \\ (I) Photoionization modelling}

\correspondingauthor{Swayamtrupta Panda}
\email{spanda@camk.edu.pl, panda@cft.edu.pl}

\author[0000-0002-5854-7426]{Swayamtrupta Panda}
\affiliation{Center for Theoretical Physics, Polish Academy of Sciences, Al. Lotnik{\'o}w 32/46, 02-668 Warsaw, Poland}
\affiliation{Nicolaus Copernicus Astronomical Center, Polish Academy of Sciences, ul. Bartycka 18, 00-716 Warsaw, Poland}

\author[0000-0002-7843-7689]{Mary Loli Mart\'inez-Aldama}
\affiliation{Center for Theoretical Physics, Polish Academy of Sciences, Al. Lotnik{\'o}w 32/46, 02-668 Warsaw, Poland}

\author[0000-0001-9719-4523]{Murilo Marinello}
\affiliation{Laborat\'orio Nacional de Astrof\'isica, R. dos Estados Unidos, 154 - Na\c{c}\~oes, Itajub\'a - MG, 37504-364, Brazil}

\author[0000-0001-5848-4333]{Bo\.zena Czerny}
\affiliation{Center for Theoretical Physics, Polish Academy of Sciences, Al. Lotnik{\'o}w 32/46, 02-668 Warsaw, Poland}

\author[0000-0002-6058-4912]{Paola Marziani}
\affiliation{INAF-Astronomical Observatory of Padova, Vicolo dell'Osservatorio, 5, 35122 Padova PD, Italy}

\author[0000-0001-5756-8842]{Deborah Dultzin}
\affiliation{Universidad Nacional Auton\'oma de M\'exico Instituto de Astronom\'ia: Ciudad de Mexico, Distrito Federal, MX 04510, Mexico}

\begin{abstract}
Optical \feii{} emission is a strong feature in quasar spectra originating in the broad-line region (BLR). The difficulty in understanding the complex \feii{} pseudo-continuum has led us to search for other reliable, simpler ionic species such as \caii{}. In this first part of the series, we confirm the strong correlation between the strengths of two emission features, the optical \feii{} and the NIR \caii{}, both from observations and photoionization modelling. With the inclusion of an up-to-date compilation of observations with both optical \feii{} and NIR \caii{} measurements, we span a wider and more extended parameter space and confirm the common origin of these two spectral features with our photoionization models using \textmyfont{CLOUDY}. Taking into account the effect of dust into our modelling, we constrain the BLR parameter space (primarily, in terms of the ionization parameter and local cloud density) as a function of the strengths of \feii{} and \caii{} emission.

\end{abstract}


\section{Introduction} \label{sec:intro}
\feii{} emission in active galactic nuclei (AGNs) covers a wide range of energy bands -- typically spreading from the ultraviolet (UV) to the near-infrared region (NIR) -- and contains numerous multiplets. These multiplets form a pseudo-continuum owing to the blending of over 344,000 transitions \citep{bv08}. Understanding the origin of the Fe emission in AGNs is a major challenge and is important for a number of reasons as summarized in \citet{murilo2016}. One of the primary reasons is its role as a strong contaminant due to the large number of lines. This may lead to an inaccurate estimate of parameters for lines from other ionic species and hence to an improper description of the physical conditions in the broad-line region (BLR). The \feii{} pseudo-continuum requires proper modelling taking into account the radiative mechanisms responsible for its emission and the chemical composition of the BLR clouds  \citep{verner99,bv08,kovacevic2010,panda18b}. 

\citet{bg92} provided one of the first \feii{} templates that has been ubiquitously used to subtract the contamination due to \feii{} in the optical band of quasar spectra. The \feii{} template has been obtained in a very simple way  by removing lines that are not \feii{} in the  PG 0050+124 (\textmyfont{I Zw 1}) spectrum. The result of their procedure was a spectrum representing only permitted \feii{} emission in \textmyfont{I Zw 1} (after removing Balmer lines, [OIII] $\lambda\lambda$4959, 5007 lines, [NII] $\lambda$5755, blend of Na I D and He I $\lambda$5876, two intense [\feii{}] lines at $\lambda$5158\AA\, and $\lambda$5273\AA). The use of I Zw 1 was motivated by the fact that emission lines in this source were very narrow and their subtraction did not affect much a broad wavelength range.

The term ``\feii{} pseudo-continuum'' drew importance after the seminal work of \citet{verner99} which developed a theoretical model for the \feii{} emission taking into account transition probabilities and collision strengths from the up-to-date atomic data available at that time (see also \citealt{sigut2003}). 
%The
Their \feii{} atom model included 371 levels (up to 11.6 eV) and predicted intensities of 68,635 lines. Later works have tested the \feii{} templates procedure by semi-empirical methods -- in the optical \citep{veron2004,kovacevic2010} and in the NIR \citep{rissmann2012}, as well as from the purely observational spectral fitting in the UV \citep{vestegaard01}.

\feii{} emission also bears extreme importance in the context of the main sequence of quasars. Several noteworthy works have established the prominence of the strength of the optical \feii{} emission (4434-4684 \AA~) with respect to the \textit{broad} \hb{} line width (henceforth \rfe{}) and it's relevance to the Eigenvector 1 sequence  primarily linked to the Eddington ratio \citep{bg92, sul00, sul01, sh14, mar18}. Recent studies have addressed the importance of the \feii{} emission and its connection with the Eddington ratio, the black hole mass, cloud density, metallicity and turbulence \citep{panda18b}, as well as with the shape of the ionizing continuum \citep{panda19}, and including the effect of the orientation of the disk plane with respect to the observer \citep{panda19b,panda19c}. 

The difficulty in understanding the \feii{} emission has motivated us to search of other reliable, simpler ionic species such as \caii{} and O I \citep[][and references therein]{martinez-aldamaetal15} which would originate from the same part of the BLR and could play a similar role in quasar main sequence studies. Here, by \caii{} emission we refer to the \caii{} IR triplet (\cat{}), consisting of lines at $\lambda$8498\AA, $\lambda$8542\AA\, and $\lambda$8662\AA. Photoionization models performed by \citet{joly1989} have shown that the relation between the ratios CaT/H$\beta$ and \rfe{} provides evidence for a common origin for the NIR \cat{} and optical \feii{}. CaT/H$\beta$ increases at high density and low temperature as does \rfe{} \citep{jol87,1999ASPC..175..303D}. \citet{persson1988} conducted the first survey for \caii{} emission in 40 AGNs. Data from \citet{persson1988} and photoionization calculations from \citet{joly1989} found that \cat{} is emitted by gas at low temperature (8000 K), high density ($>$ 10$^{11}$ cm$^{-3}$) similar to optical \feii{}. \citet{matsuoka07, matsuoka08} computed photoionization models using the O I $\lambda$8446 and $\lambda$11287 lines and \cat{}, and found that a high density ($\sim$ 10$^{11}$ cm$^{-3}$) and low ionization parameter (log U $\sim$ -2.5) are needed to reproduce flux ratios consistent with the physical conditions expected for optical \feii{} emission. 

A recent study by \citet{martinez-aldamaetal15} found the best fit relation for the CaT strength (or \cat{}/\hb{}) to that of \rfe{} is given by the following relation:
\begin{equation}
\log \left(\frac{\textmyfont{CaT}}{\textmyfont{H}\beta}\right) = (1.33 \pm 0.23) \log \left(\frac{\textmyfont{Fe II}}{\textmyfont{H}\beta}\right) - (0.63 \pm 0.07)    
\label{eq0}
\end{equation}

In this paper we reaffirm the relation between the ratios \cat{}/\hb{} (hereafter, \rcat{}) and \rfe{} using photoionization modelling and comparing with an up-to-date compilation of sources, and identify the physical conditions characterizing the emitting zones in terms of two key parameters -- \textit{ionization parameter} (U) and \textit{local cloud density} (n$_{\rm{H}}$). Section \ref{sec:methods_data} presents the sample compiled from prior observational studies, and Section \ref{sec:methods_cloudy} describes the \textmyfont{CLOUDY} \citep{f17} photoionization modelling setup, including a dust prescription applied in the post-photoionization stage of the computations. We analyse the results from the modelling and compare them with the results from the observations in Section \ref{sec:results}. We discuss certain relevant issues in Section \ref{sec:discussions} and summarize our results in Section \ref{sec:conclusions} with possible extensions in the future. Throughout this work, we assume a standard cosmological model with $\Omega_{\Lambda}$ = 0.7, $\Omega_{m}$ = 0.3, and H$_0$ = 70 \kms{} Mpc$^{-1}$.

\section{Methods} \label{sec:methods}

\subsection{Observational data} \label{sec:methods_data}

The studies of the NIR calcium triplet properties have by now some history. The first observations of this ion were published in the late-1990s \citep{persson1988}, and soon after some theoretical analysis \citep{joly1989,perssonferland89} followed. 
Up to now, 75 \cat{} measurements are available in the literature \citep{persson1988,  matsuoka2005, riffel2006, matsuoka07, martinez-aldamaetal15, martinez-aldamaetal15b, murilo2016, murilo2020}. However, not all of them are reliable.   Observations of \cat{} pre-dating NIR spectrographs on large aperture telescopes relied on optical observations. Although it is possible to observe 0.8--0.9\,$\mu$m region, where the \cat{} is located, with standard optical telescopes the observations are limited to low redshift sources. Besides, since the \cat{} equivalent width is considerably lower than for the hydrogen lines, very long exposure times are required to achieve a good signal to noise measurement of those lines. This restricts observation to bright sources and implies strong contamination of OH telluric lines present in this region. Moreover, the host  galaxy (particularly the calcium absorption lines) can be present and suppress the contribution of the \cat{} emission. Since NIR spectrographs became an option, the number of AGNs with reliable \cat{} measurements has increased, although no large survey has been carried out yet.

Optical measurements for \hb\ and \rfe\ are required in our analysis, in addition to reliable measurements of the \cat{}. As a consequence, we were able to select {58} objects. The full sample includes sources with $42.5<\mathrm{log}\,L_{5100}<46.8$ at $0.01<z<1.68$. A description of the sample and the relevant measurements for the present analysis are shown in Table~\ref{tab:table1}. The measurements are collected from the original works described in detail in the following sub-sections.  \citet{persson1988} and \citet{murilo2016} samples have a coincidence in 5 sources, and we use both measurements. These five objects are marked in Table~\ref{tab:table1}. 

\subsubsection{\citet{persson1988} sample}
The \citet{persson1988} study was the first analysis devoted to the \caii{} triplet measurements. Due to the selection criteria of the sample, the majority of the sources are mostly catalogued as NLS1 objects. Since \cat{} and \feii\ intensities are correlated, only strong \feii\ emitters were selected for ensuring the presence of the \cat{}. Also, in order to avoid the blending between \oi\ and the first two lines of the \caii{} triplet at 8498\AA\ and 8542\AA, only narrow \hb\ profiles were selected. These criteria introduce a selection effect in the sample. The original sample included 40 sources with $-27.0<M_V<-20.0$ at $0.009<z<0.159$. \citet{persson1988} divided the sample into four categories by a visual inspection considering the quality of the observation, the \cat{} strength and the presence of the host galaxy contamination. The first category includes objects without any doubt of the presence of \cat{} emission; however, some of these objects show a central dip in the \cat{} emission lines indicating the contribution from the host galaxy. The rest of the categories include sources with moderate \caii{} emission or a clear host galaxy contamination. Due to the limitation of the signal-to-noise ratio (S/N) of the sample, the decomposition of the host galaxy was not performed, and, Persson in his paper excluded 15 of the most contaminated sources.
%excluding 15 objects from the original analysis by Persson in his paper. 
For the remaining objects, the \cat{} measurement was still possible, although the reliability of the measurement had to be assessed on a case-by-case basis. We have considered the remaining 25 objects with a \cat{} measurement reported, where at least five do not show a dip in the \cat{} emission lines suggesting no contamination from a stellar continuum. Since some measurements might be problematic and others are reported as upper limits, we include the corresponding information in Table \ref{tab:table1}, where we also report the identification of the source, redshift, \rfe{}, and \rcat{}\ intensity ratio for the sources considered in this work.

\subsubsection{\citet{martinez-aldamaetal15,martinez-aldamaetal15b} sample}

With the purpose of increasing the number of objects, enlarge the redshift and luminosity range covered by the \citet{persson1988}, \citet{martinez-aldamaetal15, martinez-aldamaetal15b} observed a total 21 QSO with the Very Large Telescope (VLT) using the Infrared Spectrometer And Array Camera (ISAAC\footnote{ISAAC was decommissioned in 2013.}). The selection criteria, data reduction, and analysis are the same in the two papers of Martinez-Aldama and collaborators. Since the \cat{} has a low equivalent width compared to the equivalent width of \hb\ \citep{joly1989}, they selected the brightest (and hence most luminous, $-29<M_V<-26$) sources of the intermediate redshift QSO sample ($0.85<z<1.68$) which were later studied by \citet{sulenticetal17}. This increases the probability to observe a high signal-to-noise (S/N) spectrum in the NIR, which is required for a careful analysis. Therefore, this sample is not as biased in terms of \feii\ intensity as \citet{persson1988}'s. The majority of the sources in this sample show broad profiles, which differentiate them from the Persson and Marinello samples. While in these samples \citep[][the latter are described in the following subsection.]{persson1988,murilo2016,murilo2020} the \cat{} can be resolved individually, in Mart{\'i}nez-Aldama et al. sample the \cat{} is blended with the \oi. 

On the other hand, since that sample is of high luminosity (\lbol$\sim10^{47}$ \ergs), it is expected that the host galaxy contamination should be much smaller. In order to estimate and subtract the host galaxy contribution, they obtained the specific flux at 9000 \AA\ from the stellar population synthesis models, and then estimated the host mass assuming the $M\mathrm{_{bulge}}$/\mbh\
ratio \citep{merloni2010, magorrian1998} appropriate for the redshift of each quasar. They found that in the majority of the sample the contribution of the host is $\lesssim 10\%$, except in the quasar HE~2202-2557 where it has a contribution of $\sim50\%$ with respect to the total luminosity. In fact, before the host galaxy subtraction, the absorption features could be observed by eye in this object. Hence, the host galaxy contribution was only subtracted in HE~2202-2557.

\subsubsection{\citet{murilo2016} sample + PHL1092}
The original sample from \citet{murilo2016} consists of 25 AGNs which were selected primarily from the list of \citet{1991A&A...242...49J}. The 25 AGN spectra were taken from the AGN NIR atlas presented by \citet{riffel2016} and complemented with the spectra observed by \citet[][]{alberto2002}. Their sample was observed with SpeX \citep{rayner2003} at the 3\,m \textit{NASA Infrared Telescope Facility} (IRTF), a NIR spectrograph used in the cross-dispersed mode to observe simultaneously 0.7--2.4\,$\mu$m, with a resolution R$\sim$1300.
The additional selection criteria imposed in the NIR sample selection were: (1) the target K-band magnitude brighter than 12 to balance between a good S/N and exposure time; and (2) FWHM of the broad \hb{} component less than 3000 \kms{}. The latter limitation was introduced to ensure that severe blending from \feii{} lines with adjacent permitted and forbidden lines is avoided. Among these 25 AGNs, only 13 had \caii{} $\lambda$8662 fluxes and FWHMs available, and among these 13, only 10 had concomitant \rfe{} estimates. Hence, we consider these 10 sources in our sample. 

Besides those 10 AGNs from \citet{murilo2016}, we add to our sample an extreme-strong Fe\,{\sc ii} emitter, PHL\,1092, from \citet{murilo2020}. PHL\,1092 was the strongest Fe\,{\sc ii} emitter reported by \citet{1991A&A...242...49J}, with a \rfe{}=6.2. \citet{murilo2020} presented a panchromatic analysis of the physical properties of this AGN using a combined spectrum of HST/STIS (in the ultraviolet), SOAR/Goodman (in the optical), and GNIRS/Gemini (in the NIR), covering a wavelength range of 0.1--1.6\,$\mu$m. The high S/N and resolution from Goodman spectrograph allowed them to re-estimate \rfe{}, obtaining a value of 2.58, which still places PHL\,1092 among the strongest Fe\,{\sc ii} emitters observed.

For 4 out of 10 sources from \citet{murilo2016} sample -- 1H1934, Mrk335, Mrk493, Tons180; and for PHL1092, we notice that the emission line profiles were better represented by a Lorentzian function rather a Gaussian, as originally assumed by the authors. These sources are known narrow line Seyfert 1 galaxies, and part of the Population A \citep{Mar01}, where broad lines usually have a Lorentzian-shaped profile. We re-fit the Ca\,{\sc ii}, H$\beta$, and O\,{\sc i} lines, and re-estimate the Fe\,{\sc ii} intensity for these sources. Note that since these lines are blended -- either with its own narrow component (in the case of H$\beta$), or with other broad lines (for Ca\,{\sc ii}$\lambda\lambda8496,8542$ and O\,{\sc i}), this re-analysis results in higher fluxes of these lines, on average by $\sim$15$\%$ for the 5 sources.

There is no report of host galaxy stellar population (SP) in the sample of \citet{murilo2016} and PHL\,1092. Their sample consists of strong emission line AGNs, with low or no SP signatures in the spectra. The two main SP features in the NIR spectrum are the CO bands redwards of 2.3$\mu$m and the \caii{} triplet absorption around 0.8$\mu$m \citep{riffel2006}. Since their sample is dominated by emission lines, the absorption around \caii{} triplet is not visible in the spectra. Moreover, the redshifts of the sources do not allow a reliable measurement of the CO bands, since it is outside the wavelength coverage of the instrument. For these reasons, \citet{murilo2016} considered an error of up to 10\% due to the host galaxy on the continua in their measurements.

%%%%%%%%%%%%%%%%%%%%%%%%%%%%%%%%%%

\subsection{Photoionization modelling setup} \label{sec:methods_cloudy}
We perform a suite of \textmyfont{CLOUDY} models aimed at understanding the physical conditions in the BLR medium leading to efficient production of \cat{} and to test the connection between \cat{} and Fe\,{\sc ii} emissivities.  Our simulated grid was created by varying the cloud particle density, $10^{5} \leq \rm{n_H} \leq 10^{13}\;(\rm{cm^{-3}}$), and the ionization parameter, $-7 \leq \log \rm{U} \leq 0$. The modelling setup applies a single cloud model, but assumes a grid of ionization parameters and cloud densities, which allows to reproduce a range of BLR radii for a fixed number of ionizing photons emanating from the central region. We do not apply the more complex \textit{Locally Optimized Cloud} (LOC) prescription \citep{baldwin1995} which requires integration of the cloud emissivity over a range of densities and radii that are assumed to have power law distributions for each model. Such a model has more parameters since, apart from the power law slopes, the results do depend on the adopted outer radius of the BLR and the maximum allowed cloud density \citep[see e.g.][]{goad2015}. The single cloud model, although simplistic in its approach, has proved to be quite successful in estimating the optical \feii{} emission line strengths as has been verified in our previous works \citep{panda_frontiers, panda18b, panda19, panda19b, panda19c}. It was also used by \citet{negreteetal13} where they were able to reproduce three line ratios without restoring to LOC model. There is a theoretical argument in favor of the universal local density of the BLR clouds which is based on radiation pressure confinement \citep{baskin2018}. The remaining parameters are the cloud column density ($N_H$) for which we assume a value of $10^{24}\;\rm{cm^{-2}}$ in our Base Model, motivated by our past studies \citep{panda_frontiers,panda18b,panda19}. Such a column density is sufficiently large to give low ionization lines while still keeping the medium optically thin. This allows us to neglect the additional effect from the electron scattering that starts to play a role for media that are optically thick ($\tau \gtrsim 10$). In our Base Model we also fix chemical composition of the medium at solar abundance (\zsun{}) which are estimated using the \textit{GASS10} module \citep{gass10}. The effect of the change in metallicity is addressed in Section \ref{sec:metallicity} accounting for a sub-solar case (Z = 0.2\zsun{}) and a super-solar case (Z = 5\zsun{}). We also test the effect of the change in column density to higher values, such as, $10^{24.5}\;\rm{cm^{-2}}$ and $10^{25}\;\rm{cm^{-2}}$, which is addressed in Section \ref{sec:column_density}. Indeed, one expects a broad range of column densities to be present in the BLR, yet, this value of the $N_H$ quite consistently reproduces the observed line emission, especially in the case of the optical and UV \feii{} as shown in \citet{bv08}. We utilize the spectral energy distribution (SED) for the nearby (z= 0.061) Narrow Line Seyfert 1 (NLS1), \textmyfont{I Zw 1}\footnote{The \textmyfont{I Zw 1} ionizing continuum shape is obtained from \href{http://vizier.u-strasbg.fr/vizier/sed/}{Vizier photometric viewer}.}. We assume that the observed SED is also the one seen by the BLR gas. The parameters \rfe\ and \rcat{} are extracted from these simulations\footnote{The \textmyfont{CLOUDY} output files for each model (inclusive of the SED shape and a sample input) are hosted on our GitHub repository: \href{https://github.com/Swayamtrupta/CaT-FeII-emission}{https://github.com/Swayamtrupta/CaT-FeII-emission}.}, where \rfe{} is the \feii{} intensities integrated between 4434-4684 \AA~ and normalized by the broad \hb{} $\lambda$4861\AA~ intensity, and similarly, \rcat{} is the sum of the intensities for the \caii{} triplet at $\lambda$8498\AA, $\lambda$8542\AA\, and $\lambda$8662\AA\, normalized by the same broad \hb{} intensity.

\subsection{Inclusion of dust sublimation} \label{sec:methods_dust}

\begin{figure*}%
    \centering
    \includegraphics[width=2.1\columnwidth]{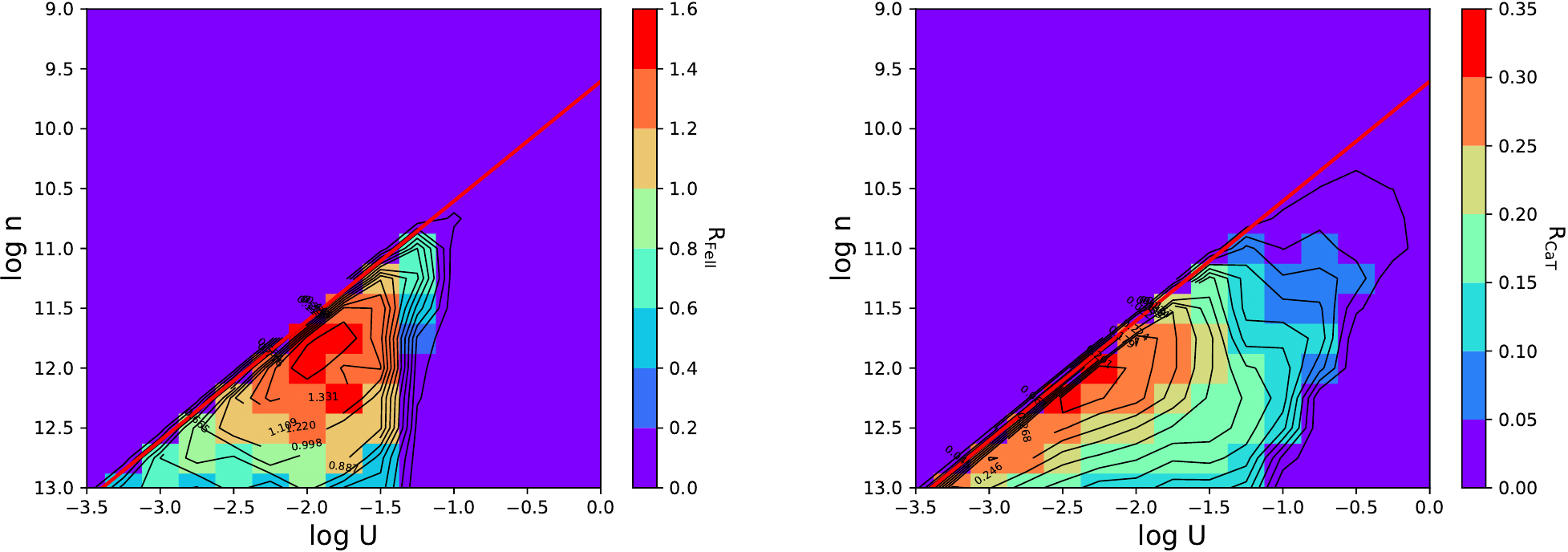}
    \caption{2D histograms showing log U - log $\rm{n_H}$ for (a) \rfe{}; and (b) \rcat{}; for the base model, i.e. at Z = \zsun{}, N$_{\rm{H}}$ = 10$^{24}\;$cm$^{-2}$. The solid red line denotes the dust sublimation radius ($R_{sub} = 0.83\;pc$) considering \citet{Nenkova2008} prescription. The black contours also represent the \rfe{} and \rcat{} emission, respectively.}%
    \label{fig:cat_rfe2}%
\end{figure*}

We assume that the BLR outer radius is given by the inner radius of the dusty/molecular torus \citep{netzer1993} which constrains the available parameter space. Our approach assumes dustless clouds in the BLR, and the dust sublimation temperature limits the considered size of the BLR. As shown in \citet{kishimoto07} (adopted from \citealt{barvainis1987}), the dust sublimation radius depends collectively on the source luminosity (L$_{\rm{UV}}$, coming from the accretion disk), the sublimation temperature of the dust (T$_{\rm{sub}}$), and the dust grain size (a),
\begin{equation}
    \rm{R_{sub}\;[pc] = 1.3\left(\frac{L_{UV}}{10^{46}}\right)^{0.5}\left(\frac{T_{sub}}{1500\;K}\right)^{-2.8}\left(\frac{a}{0.05\;\mu m}\right)^{-0.5}}
    \label{eq1}
\end{equation}
The prescription from \citet{Nenkova2008} which we apply to estimate the dust sublimation radius in this paper assumes, for simplicity, a dust temperature $T_{sub}$ = 1500 K, which has been found consistent with the adopted mixture of the silicate and graphite dust grains. The actual form of their Equation 1 is given as:
    \begin{equation}
      \rm{R_{sub}\;[pc] = 0.4\left(\frac{L_{UV}}{10^{45}}\right)^{0.5}\left(\frac{T_{sub}}{1500\;K}\right)^{-2.6}}  
      \label{eq2}
    \end{equation}
This has a slightly different form than Equation \ref{eq1} which is originally derived in \citet{barvainis1987}. \citet{Nenkova2008} solves the radiative transfer problem for smooth density distributions which does not involve separately the size of the dust grains or their volume density, and hence the grain size parameter is dropped (for a= 0.05 $\mu$m). This is the origin for the slightly steeper exponent for the dust-temperature part in their relation compared to \citet{barvainis1987}. Upon simplification, the prescription from \citet{Nenkova2008} has a form: $\rm{R_{sub} = 0.4\left(L_{UV}/10^{45}\right)^{0.5}}$ [parsecs], where, $\rm{R_{sub}}$ is the sublimation radius computed from the source luminosity that is consistent for a characteristic dust temperature. The sublimation radius is estimated using the integrated optical-UV luminosity for \textmyfont{I Zw 1}. This optical-UV luminosity is the manifestation for an accretion disk emission and can be used as an approximate of the source's bolometric luminosity. The bolometric luminosity of \textmyfont{I Zw 1} is L$_{\rm{bol}} \sim 4.32\times10^{45}$ erg s$^{-1}$. This is obtained by applying the bolometric correction prescription from \citet{netzer2019} to \textmyfont{I Zw 1}'s optical monochromatic luminosity, $\rm{L_{5100}} \sim 3.48\times 10^{44}\;\rm{erg\;s^{-1}}$ \citep{persson1988}. This uniquely sets the dust sublimation radius at $\sim 0.83 \rm{pc}\;( = 2.56\times 10^{18}\;$cm). Projecting this sublimation radius on the \un{} plane allows us to recover the non-dusty region that well represents the physical parameter space consistent with the emission from the BLR (see Fig. \ref{fig:cat_rfe2}). This dust-filtering is applied to the models in a post-photoionization stage.

\section{Analysis and Results} \label{sec:results}

\subsection{Physical conditions in the {\sc CaT} and {\sc Fe\, ii} emitting regions}

We used our basic simulation grid to compare the parameter ranges of density and ionization parameter corresponding to the most efficient production of \cat{} and \feii{}. The left panel of Figure \ref{fig:cat_rfe2} shows the log U - log $\rm{n_H}$ plane as a function of the ratio \rfe{}; the right panel shows the corresponding distribution for \rcat{}. The red line limits the region to the dustless BLR\footnote{Note: the limits on the colorbar are different for the left and right panels. This is done to emphasize the zone of maximum emission in both these cases. As the two ratios are normalized with respect to the intensity of H$\beta$, we effectively map the emitting regions for the \feii{} and \cat{}, respectively.}. Clearly, the zones are very similar, although the \cat{} emission is seemingly more extended than the one for \feii{} -- stretching to densities $\lesssim$ 10$^{9.5} \;\rm{cm^{-3}}$ albeit at very high ionization (log U $\gtrsim -1$). Also, the maximum of the emission for \cat{} is concentrated in a region which has relatively high densities ($10^{11.5} \lesssim \rm{n_H} \lesssim 10^{12.25}\;\rm{cm^{-3}}$) but requires lower ionization parameters ($-2.5 \lesssim \log \rm{U} \lesssim -2.0$) than \feii{}. Furthermore, we find that the maximum \rfe{} is by a factor $\sim$4.5 higher than the corresponding \rcat{} maximum.

\begin{figure}[h]
\centering
\includegraphics[width=.5\textwidth]{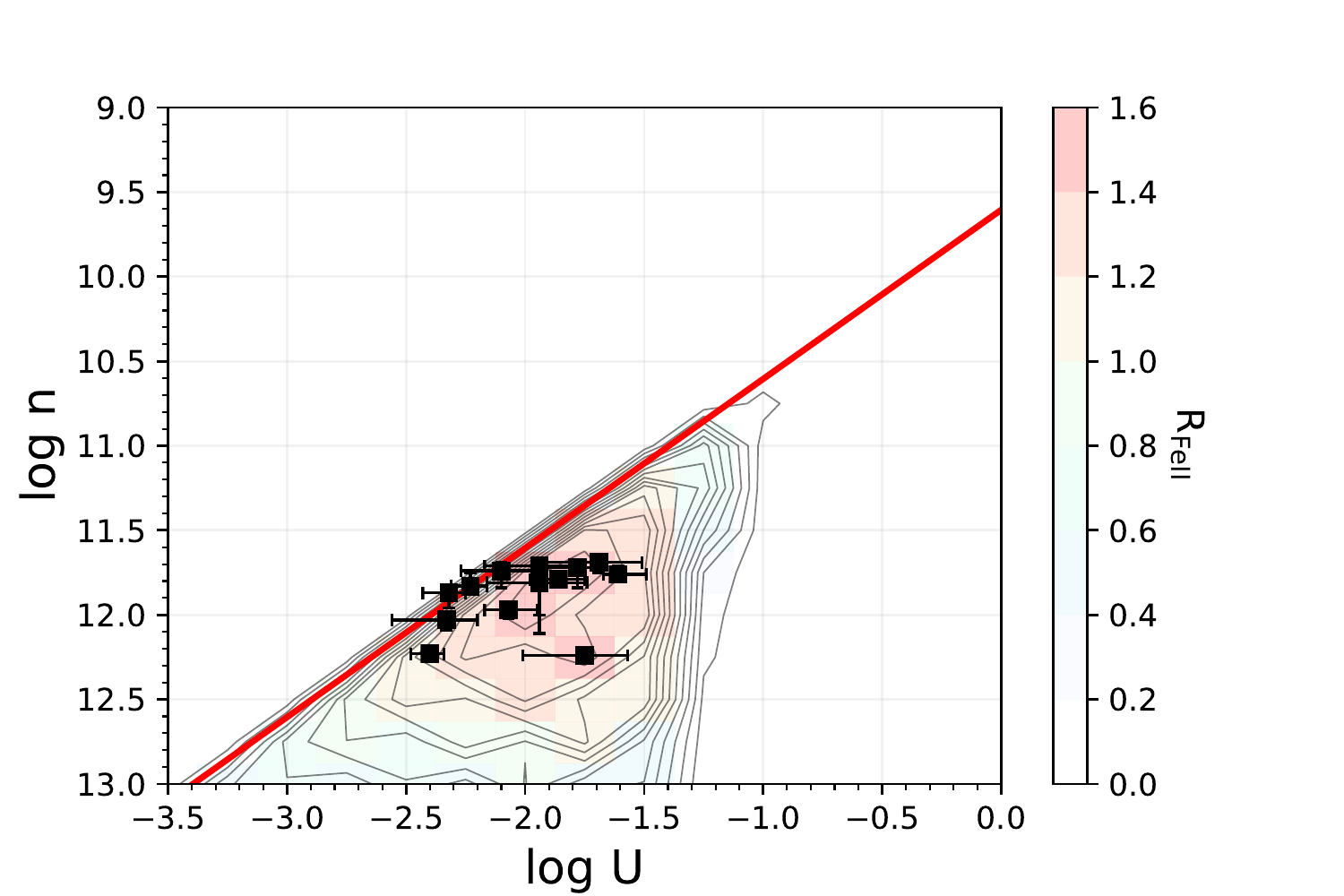}
\includegraphics[width=.5\textwidth]{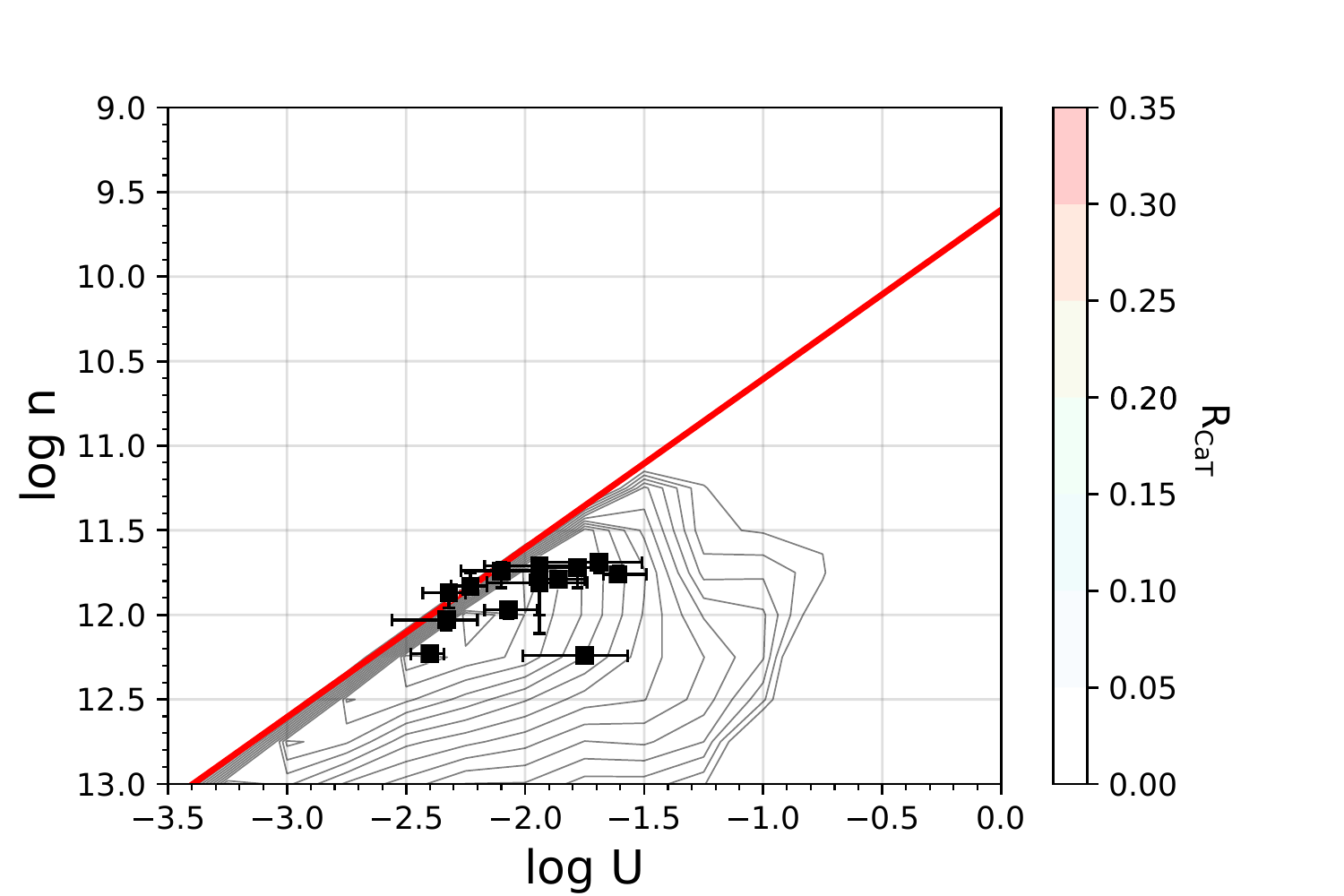}
\caption{\un{} estimates for 13 \textit{reverberation-mapped} AGNs from \citet{negreteetal13}. Our photoionization setup color-coded with \rfe{} (upper-panel) and with \rcat{} (lower-panel) is shown in the background (identical to Figure \ref{fig:cat_rfe2}).}
\label{fig2:fe2_in_blr}
\end{figure}

A related fundamental question in this context is -- How to confirm whether the \feii{} emission is confined to the BLR? Previous studies \citep[][ and references therein]{panda18b} have been able to quantify the \feii{} emission in the BLR and compare it to the values obtained from observations -- using the DR7 quasar catalogue \citep{s11}. They further constrained the location of the emitting region for the species by studying their emissivity profiles with respect to \hb{}. The emissivity profiles provide a confirmation of the results obtained from the time-lag measurements of \feii{} based on reverberation mapping of selected AGNs \citep{2013ApJ...769..128B, hu15}. To confirm this in the context of the current study, we procure a sample of reverberation-mapped selected AGNs from \citet{negreteetal13} wherein the authors had estimated the size of the BLR using the photoionization method by employing diagnostic line ratios in the UV. In addition to confirming the BLR size which were found in good agreement to the reverberation-mapped estimates, the authors were able to estimate the ionization parameter and the density for the 13 sources. Figure \ref{fig2:fe2_in_blr} shows these estimates in the \un{} space overlaid on top of our photoionization model setup for both \rfe{} and \rcat{} cases. This further confirms that both these emitting regions are present in the BLR considered in this work.
The observational points from that sample may imply that the actual parameter range in BLR is narrower than adopted in our simulations, but the sources populate the region of the highest emissivity.

\subsection{\rcat{} vs. \rfe{}: observations}

Figure \ref{fig:ratio_base_model} shows the correlation between the \rcat{} and \rfe{} both from observations and from our photoionization modelling (Base model). For the observed sample the correlation between the \rcat{} and \rfe{} is strong and significant, with the correlation coefficient \textit{r} $\approx$ 0.737, and a p-value = 4.33$\times 10^{-11}$. The best-fit relation for the full sample from observations is:
\begin{equation}
\begin{aligned}
\log \left(\frac{\textmyfont{CaT}}{\textmyfont{H}\beta}\right) \approx & (0.974 \pm 0.119) \log \left(\frac{\textmyfont{Fe II}}{\textmyfont{H}\beta}\right) \\
& - (0.657 \pm 0.041)
\end{aligned}
    \label{eq:data}
\end{equation}
which has a slope consistent with simple proportionality between the two quantities, with more scatter in the lower range. The best fit slope reported by \citet{martinez-aldamaetal15} was considerably higher (1.33 vs. 0.974 in this paper) but the two values are still consistent within a 3$\sigma$ limit. The change in the slope resulted from inclusion of more data, particularly half of \citet{murilo2016} sample populate the bottom-right part of the relation resulting in a slightly shallow trend. In fact, this sample includes objects with the highest \rfe\ and \rcat{} values, which were underrepresented in the previous sample of \citep{martinez-aldamaetal15, martinez-aldamaetal15b}. In an upcoming paper, we will explore the different kinds of AGN included in our sample considering the 4DE1 scheme \citep{sulentic2000, marzianietal18}.

We decided to keep also the results from the older observations \citep{persson1988} which were re-observed and re-analysed in \citet{murilo2016}. The more recent observations of the  five sources that are in common show a larger \rfe{} and \rcat{}. These differences can be caused by different methods of fitting the Fe\,{\sc ii} bump and the H$\beta$ profile. For instance, the work of \citet{persson1988} is prior to \citet{bg92} and thereby does not use the same Fe\,{\sc ii} template as \citet{murilo2016}. Also, the analysis of the spectra for sources observed other than the \citet{persson1988} use Lorentzian profiles \citep[e.g.,][]{sulenticetal02,craccoetal16,negreteetal18} to fit the H$\beta$ profile that is typically considered for NLS1s. This additionally introduces a bias in the fluxes and FWHM measured, increasing the values of the ratios compared to the \citet{persson1988} sample.

\subsection{\rcat{} vs. \rfe{}: photoionization modelling}
\label{sec3.3}

\begin{figure*}
    \centering
    \includegraphics[width=1.65\columnwidth]{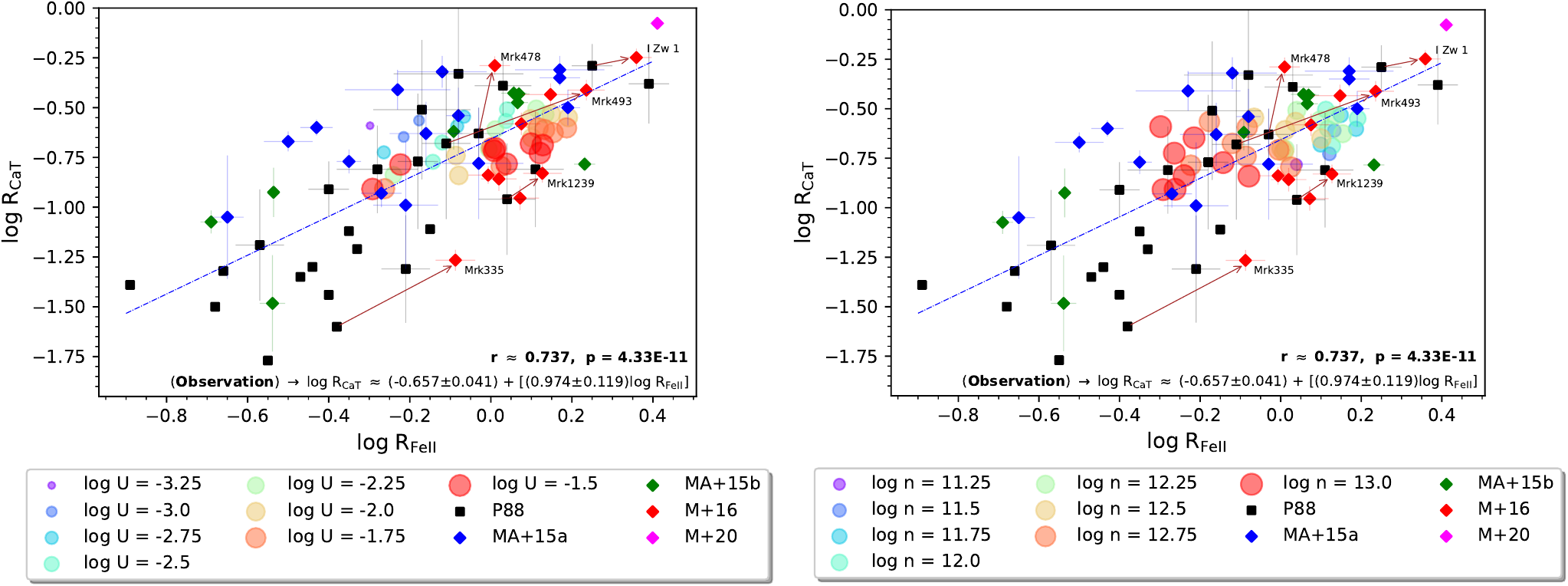}
    \caption{LEFT: Observational measurements (with errors) for \rfe{} and \rcat{} are from \citet{persson1988} (black squares), \citet{martinez-aldamaetal15} (blue diamonds), \citet{martinez-aldamaetal15b} (green diamonds), \citet{murilo2016} (red diamonds) and for PHL 1092 estimates from \citet{murilo2020} (magenta diamond). The best fit for the observational sample is shown with a dashed blue line and also reported on the plot with corresponding Pearson's correlation coefficients and null probabilities (p-values). The arrows show the shift in the \rcat{} and \rfe{} estimates from older \citep{persson1988} observations to newer, higher S/N observations \citep{murilo2016} for the five common sources with \rcat{} and \rfe{} estimates. Data points from the photoionization modelling at solar abundance (Z=\zsun{}) are shown as a function of \textit{ionization parameter (U)} in log-scale (i.e. Base Model). The column density ($\rm{N_{H}}$) for the model is assumed to be at $10^{24}$ cm$^{-2}$. RIGHT: Data points from the photoionization modelling at solar abundance (Z=\zsun{}) are shown as a function of \textit{cloud density ($\rm{n_{H}}$)} in log-scale. Rest-frame parameters are identical to the left panel.}
    \label{fig:ratio_base_model}
\end{figure*}

The results of our photoionization simulations using \textmyfont{CLOUDY} facilitates the comparison of the \rcat{} and \rfe{} to those compiled in our catalogue as described in Sec. \ref{sec:methods_data}. We extract the modelled data from the regions as shown in  Figure \ref{fig:cat_rfe2} and plot them along with the data from our observed sample. This is shown in Figure \ref{fig:ratio_base_model}. 

The two panels of Figure \ref{fig:ratio_base_model} show the correlation between the \rcat{} vs. \rfe{} both from the observations and from our base model from \textmyfont{CLOUDY}. The two figures are identical except that the left panel shows the modelled data-points color-coded (and with increasing point-size) as a function of the ionization parameter, while the right panel shows the same data as a function of the cloud density. Points representing the model cover well the central part of the observational plot, although they do not extend far enough to cover either tail of the observed distribution. Higher concentration of the theoretical points partially accounts for the disagreement with the linear slope for the best-fit to the \rcat{}-\rfe{} correlation obtained from the observational sample. %Plots show the trends with the adopted values of the parameters.
In the left panel, we see that the maximum \rfe{} is obtained for ionization parameters, $-2.0 \lesssim \log\;\u{} \lesssim -1.75$. For the maximum in \rcat{}, this value is slightly lower, $-2.75 \lesssim \log\;\u{} \lesssim -2.25$. From the modelled data, there is a clear truncation in the \rcat{} around $\log\,$\rcat{} $\approx$ -0.5. Similarly, in the right panel, we find that the cloud density required for the maximum \rfe{} is \n{} $\sim$ 10$^{12}$ cm$^{-3}$ which reconfirms the conclusions from our previous works \citep[see][are references therein]{panda19c}. This value of cloud density is also consistent with maximising the \rcat{} emission in our base model. These results are consistent with the findings from \citet{martinez-aldamaetal15} who have suggested that the \cat{} emitting region is located at the outer part of a high-density BLR. For the quoted values of \u{} and \n{}, applying the standard photoionization theory, this corresponds to a radial scale\footnote{The number of ionizing photons for \textmyfont{I Zw 1} is derived from the bolometric luminosity, L$_{\rm{bol}} \sim 4.32\times10^{45}$ erg s$^{-1}$, assuming that the net ionizing photon flux, \textit{Q(H)} = $L_{bol}/h\nu$, where the \textit{h$\nu$} $\sim$ 1 Rydberg \citep{wandel99,marz15}. See also the description of the product of ionization parameter and density for our base model in the appendix of this paper (Sec. \ref{sec:product}).}, r $\lesssim$ 0.1 pc. Although, in our base model, increasing the density above this limit (\n{} $\gtrsim$ 10$^{12}$ cm$^{-3}$) leads to a drop in both \rfe{} and \rcat{} emission similar to what was deduced from the ionization case (see also Figure \ref{fig:ratioZ1_Un}). Thus, at least for the strong \feii{} emitters, densities higher than 10$^{12}$ lead to contradictory results. We test these hypotheses in the following sections by varying the physical parameters in our \textmyfont{CLOUDY} models. 

Throughout the paper, we utilize only the intensity ratios for the two ionic species and interpret the results based on these estimates. Another important quantity is the equivalent width (EW) of the line which gives a quantitative measure of the spectral feature. In the context of this work, we find that the I(\feii{})/I(\hb) $\sim$ EW(\feii{})/EW(\hb) (where, \textit{I} denotes the line intensity). This is same for the \cat{}. We note that, in the context of photoionization, the EW can be adjusted by changing the covering fraction in the models to match the observed line widths. The problem to reproduce the EW of Low Ionization Lines has been discussed in the literature before. Either additional mechanical heating is necessary \citep[e.g.][]{collin-souffrin1986,jol87}, or multiple cloud approach, with part of the radiation scattered/re-emitted between different clouds, or BLR does not see the same continuum as the observer \citep[e.g.][]{kor97}. Previous estimates of the BLR size based on the ionization parameter gave much larger values than the reverberation measurements. All this  may not affect the line ratios as much as the line EW.

\begin{figure*}
    \centering
    \includegraphics[width=1.65\columnwidth]{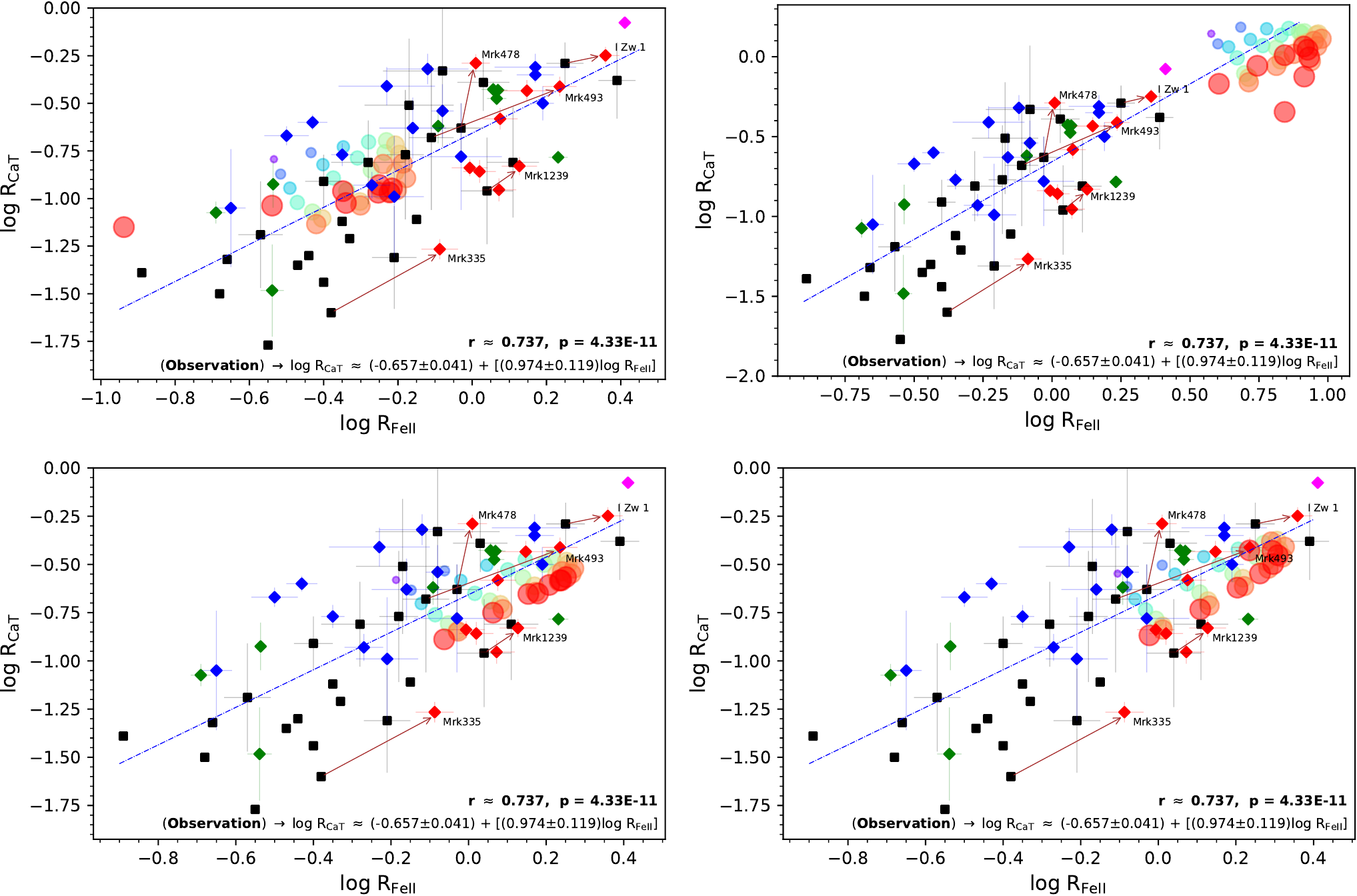}\\
    \vspace{0.1cm}
    \hbox{\hspace{5.5em}\includegraphics[width=0.825\linewidth]{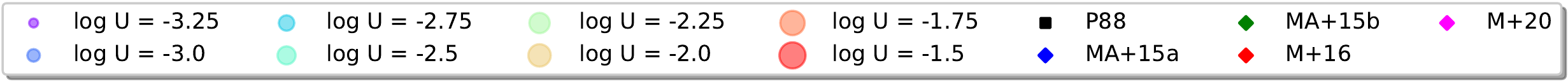}}
    \caption{TOP LEFT: Same as left panel of Figure \ref{fig:ratio_base_model} at Z=0.2\zsun{}. Data points from the \textmyfont{CLOUDY} model are shown as a function of \textit{ionization parameter (U)} (in log-scale). TOP RIGHT: Same as previous panel but at Z=5\zsun{}. BOTTOM LEFT: Same as left panel of Figure \ref{fig:ratio_base_model} but at column density, $\rm{N_{H}}$=$10^{24.5}$ cm$^{-2}$. BOTTOM RIGHT: Same as previous panel but at column density, $\rm{N_{H}}$=$10^{25}$ cm$^{-2}$.}
    \label{fig:ratioZNHvary}
\end{figure*}

\subsubsection{Effect of the metallicity} \label{sec:metallicity}

In \citet{panda18b}, we found that the increase in the metallicity in the BLR cloud leads to an increase in the net \feii{} emission. This is shown also in the earlier works from \citet{hamannferland92, leighly04} and was later extended in \citet{panda19,panda19b,panda19c} to explain the high \feii{}-emitters\footnote{Sources with \rfe{} $\gtrsim$ 1, which essentially belong to the NLS1s type with FHWM(\hb{}) $\leq$ 2000 \kms{} \citep[see][and references therein]{2020arXiv200207219M}.} in the main sequence of quasars. In the base model for the \rcat{} versus \rfe{}, the modelled objects are unable to explain the highest emitters ($\log\,$\rfe{} $\gtrsim$ 0.2) as well as the very-low emitters ($\log\,$\rfe{} $\lesssim$ -0.6). 

We investigate this effect by testing with two different cases other than the solar abundances, i.e. (a) with sub-solar case (see \citealt{punslyetal18} for the analysis of a low \feii{} emitter implying slightly sub-solar metallicity) where we assume a net metallicity of Z = 0.2\zsun{} (see top-left panel of Figure \ref{fig:ratioZNHvary}); and (b) with super-solar case where we assume a net metallicity of Z = 5\zsun{} (see top-right panel of Figure \ref{fig:ratioZNHvary}). The super-solar values of metallicities have been studied and confirmed in \citet[][and references therein]{negreteetal12,panda19b}. 

For (a) Z = 0.2\zsun{} (see top-left panel in Figure \ref{fig:ratioZNHvary}), we see that the sources that are at the lower end of the observed sample's best-fit correlation can now be explained with the considered range of \un{} values, down to $\log\,$\rfe{} $\sim$ -0.2. The model is unable to cover the lower-end (log \rcat{} $\lesssim$ -1.25) for \rcat{} emission, especially the sources from the \citet{persson1988} sample, and correspondingly for \rfe{} $\lesssim$ -0.6 (apart from one modelled data-point, i.e. log \u{} = -1.5, with log \n{} = 13.0). Defined trends almost tangential to the best-fit line can be seen which highlight the increasing \rcat{} with decreasing ionization parameter. Each of these trends corresponds to a particular value of the cloud density (see also the top-left panel of Figure \ref{fig:ratioZNHvary_nbased} where the modelled data are shown as a function of the cloud density). The super-solar case and the cases with higher column densities reiterate this trend (see Sec. \ref{sec:column_density}). For each of the segments, the peak values correspond to a $\log\,$\u{}$\cdot$\n{} = 9.75 creating an impression of truncation in the maximum \rcat{} obtained from this model.

In case (b) Z = 5\zsun{}, none of the observed sources can be modelled with this scaling in the metallicity. The entirety of the modelled data shifts to the higher \rfe{} (and correspondingly higher \rcat{}), even though the modelled data points stay on top of the observed sample's best-fit correlation.

This confirms the role of metallicity as a scaling parameter, where an increasing metallicity enhances both \cat{} and \feii{} emission. We showed here the results from two representative cases (Z = 0.2\zsun{} and Z = 5\zsun{}) apart from the base model at solar metallicity. The almost linear scaling of \rcat{} and \rfe{} values in terms of the increasing metallicity provides constraints on this parameter based on the observed sample. Thus, a metallicity value  \zsun $<$ Z $<$ 5\zsun{} can explain high \feii{} emitters in our observed sample. Although, the exact nature of the effect of the metallicity on the trends observed is still not certain. We expect that, on top of the effect of metallicity, sources like the PHL1092 -- which is the highest \feii{} emitting source in our sample and also accreting at a very high rate \citep[\LLEdd{}=1.24,][]{murilo2020}, can be modelled with an appropriate incident continuum, one that has a relatively higher number of energetic photons ($\gtrsim$1 Ryd). As has been found in recent works \citep{ferlandetal2020,panda19},
the role of the soft X-ray excess is important to extract emission, especially from the low ionization lines such as \feii{}, and the shape of the ionizing continuum is a
crucial ingredient in boosting the net estimate for \rfe{} and \rcat{} alike. This will be explored in a future work.

\subsubsection{Effect of the cloud column density}  \label{sec:column_density}

After the release of the \citet{persson1988}  sample with both \cat{} and \hb{} coverage that allowed the measurement of \rfe{} and \rcat{} for the same sources, \citet{joly1989} and \citet{perssonferland89} performed theoretical analyses using photoionization to understand the relations involving \feii{}, \hb{} and \cat{}.  \citet{perssonferland89} made a series of photoionization calculations, including heating due to free-free and H$^{-}$ absorption. These processes couple the NIR to millimeter continuum with emitting gas and are often the main agents heating the clouds at large column densities. They concluded that no matter the ranges of density and ionization parameters considered, the observed \cat{} emission cannot be modelled with relatively thin clouds, i.e. with column densities $\sim\,10^{23}$ cm$^{-2}$. They found that the gas emitting the \cat{} emission lines is basically the same that is responsible for \feii{} emission, although because of the lower ionization potential\footnote{the IPs are retrieved from the NIST Atomic database \href{https://physics.nist.gov/PhysRefData/ASD/lines\_form.html}{https://physics.nist.gov/PhysRefData/ASD/lines\_form.html}}, $\sim\,$3.1 eV \citep[see Figure 1 in][]{azevedo06} as compared to \feii{}'s $\sim\,$5.5 eV \citep[see Figure 13 in][]{shap12} the \cat{} emission arises from regions that are deeper within the neutral zone. This is also explained in the Figure 7 from \citet{perssonferland89}, where they find that for an assumed column density, $N_H = 10^{24.5}$ cm$^{-2}$ and above, the \rfe{} leads the \rcat{} by a factor of $\sim$ 1.6 with a much extended emitting region by a factor of $\sim$ 3-30. The authors have also suggested that such large column densities could be related to the presence of wind or corona above the accretion disk.

We adopt this idea by considering two additional cases with increased column densities with respect to our base model: (a) $N_H = 10^{24.5}$ cm$^{-2}$ (see lower-left panel of Figure \ref{fig:ratioZNHvary}), and (b) $N_H = 10^{25}$ cm$^{-2}$ (see lower-right panel of Figure \ref{fig:ratioZNHvary}).  Now, with the increase in $N_H$ from $10^{24}$ to $10^{24.5}$ cm$^{-2}$, the bulk of the modelled data shifts to higher values in both \rfe{} and \rcat{} (by $\sim$32\% in \rfe{} and $\sim$12\% in \rcat{}, with respect to the base model). Similar to the low-metallicity case (Z=0.2\zsun{}), we see defined trends that show increasing \rcat{} with decreasing ionisation parameter and the peak values in each trend correspond to a $\log\,$\u{}$\cdot$\n{} = 9.75 (see also the lower panels of Figure \ref{fig:ratioZNHvary_nbased}). The highest values of ionization parameters (i.e., log \u{} = -1.5) form a lower wall parallel to the best-fit relation as a function of varying densities. The data-point for the lowest \rcat{} recovered (log \rcat{} $\sim$ -0.889) for this parallel trend corresponds to the highest density (log \n{} = 13.0 cm$^{-3}$), implying that, for a fixed luminosity, the product of the ionization parameter (log \u{} = -1.5) and density (log \n{} = 13.0 cm$^{-3}$) corresponds to the smallest BLR radius\footnote{we refer the reader to \citet{2020arXiv200413113P} for more details on the estimation of the radial sizes in the context of this work.} ($\sim$0.01 pc). There's an increase in the maximum \rfe{} recovered in this model ($\log\,$\rfe{}$\sim$0.272) which is almost consistent with the \rfe{} value estimated for \textmyfont{I Zw 1} from the older observations \citep[$\approx$-0.25,][]{persson1988}.

For (b) $N_H = 10^{25}$ cm$^{-2}$, there's a further shift in the bulk of the modelled data (by $\sim$50\% in \rfe{} and $\sim$27\% in \rcat{}, with respect to the base model). The maximum \rfe{} recovered ($\log\,$\rfe{}$\sim$0.326) increases by $\sim$56\% compared to the base model. For the \rcat{}, the maximum value ($\log\,$\rcat{}$\sim$-0.374) went up by a similar factor ($\sim$53\%) although even with this model, we are unable to account for the \rfe{} and \rcat{} measurements for the highest \feii{} emitting sources in our sample, i.e., PHL1092 (shown with a magenta diamond), I Zw 1 estimate from \citet{murilo2016} and Mrk231 from \citet{persson1988}. There is a clear U-turn feature observed in the higher ionization cases (log \u{} = -1.5 and -1.75), where the maximum \rfe{} values corresponds to a singular cloud density, log \n{} = 11.75 and drops when going in either direction in terms of the density. With an additional increase in the column density ($ 10^{25.5} \lesssim N_H \lesssim 10^{26}$ cm$^{-2}$), we can extend the correlation to \rfe{} values consistent with the ones of PHL1092 and others, but then the optically-thin approximation ($\tau \sim 1$) assumed in this work is no longer applicable.

We conclude that lower column densities of the order of $10^{24}$ cm$^{-2}$ are sufficient for the low \feii{}--\caii{} emitters, yet higher column densities ($\gtrsim 10^{24.5}$ cm$^{-2}$) are required for the strong \feii{}-\caii{} emitters (\rfe{} $\gtrsim$ 1). It is evident from our analysis that fitting the observed trends requires a coupling between the column density and metallicity. Increasing the values for both these parameters allows to recover higher \rfe{} and \rcat{} estimates. On the other hand, lowering them recovers the low emitters. We explore this dependence of column density and metallicity in more detail in a subsequent work \citep{2020arXiv200413113P}.

\begin{figure}
    \centering
    \includegraphics[width=0.95\columnwidth]{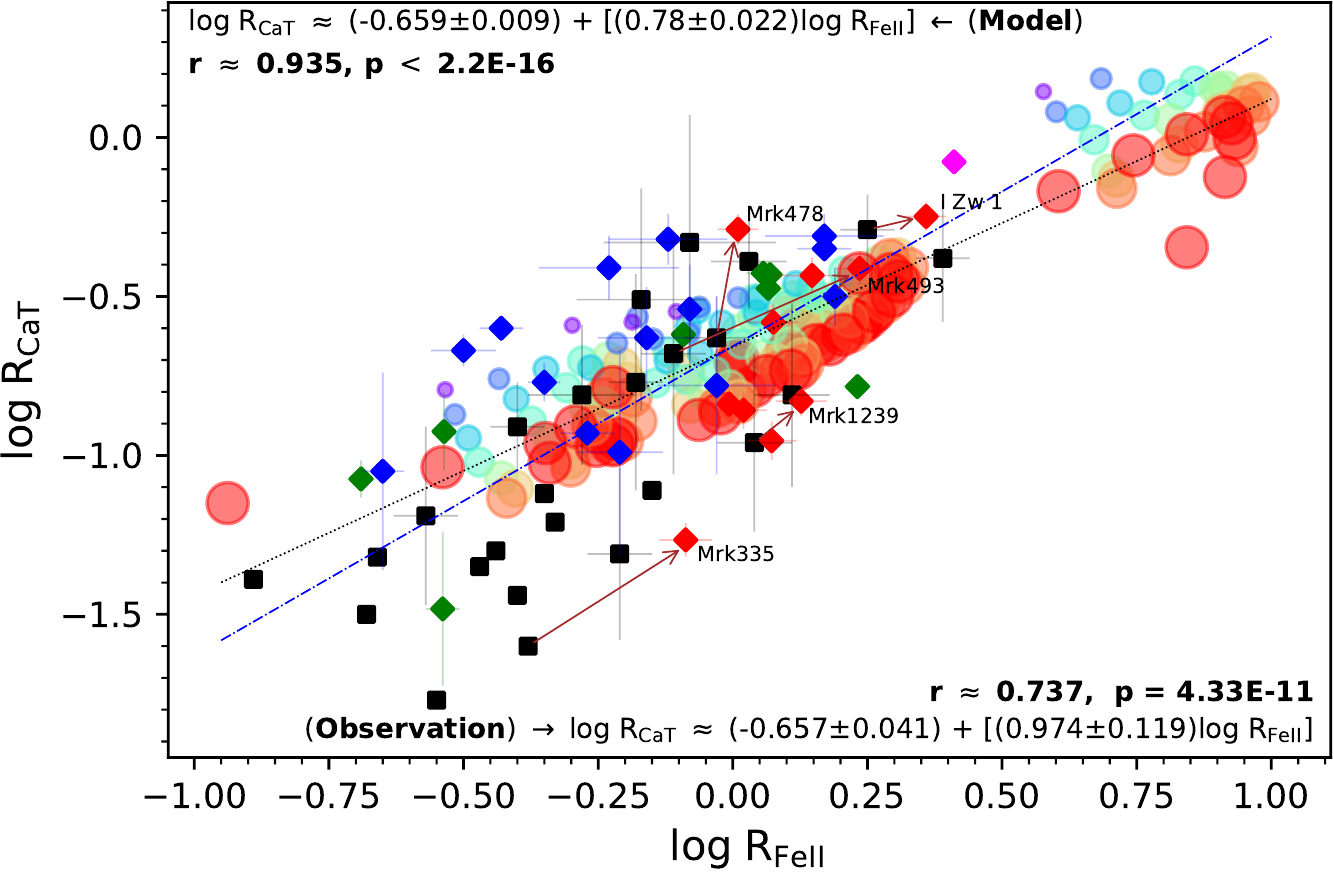}\\
    \hbox{\hspace{2em}\includegraphics[width=0.925\linewidth]{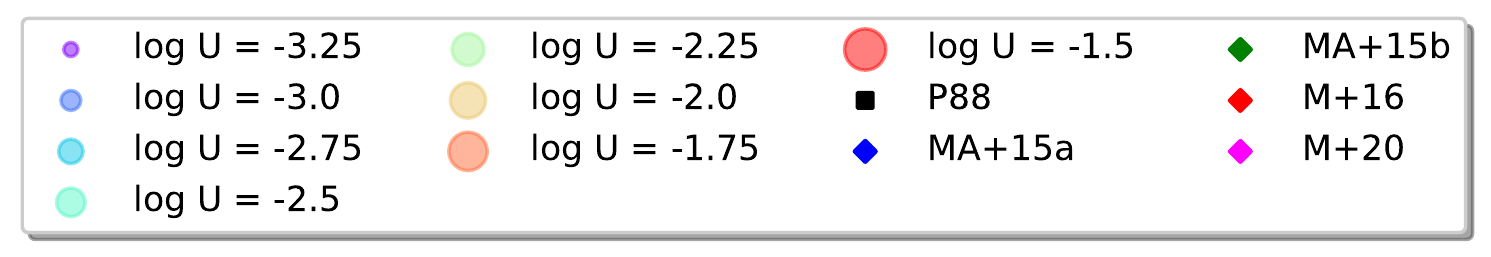}}
    \caption{Same as the left panel of Figure \ref{fig:ratio_base_model} with modelled data combined from the previous correlation plots, i.e. Figure \ref{fig:ratio_base_model} (left panel) and all panels from Figure \ref{fig:ratioZNHvary}. Data points from the \textmyfont{CLOUDY} models are shown as a function of \textit{ionization parameter (U)} in log-scale) identical to the left panel of Figure \ref{fig:ratio_base_model}. The best fits for the observational sample (dashed blue) and for the modelled data points (dotted black) and also reported on the plot with corresponding Pearson's correlation coefficients and null probabilities (p-values).}
    \label{fig:combined}
\end{figure}

\begin{figure*}
    \centering
    \includegraphics[width=1.65\columnwidth]{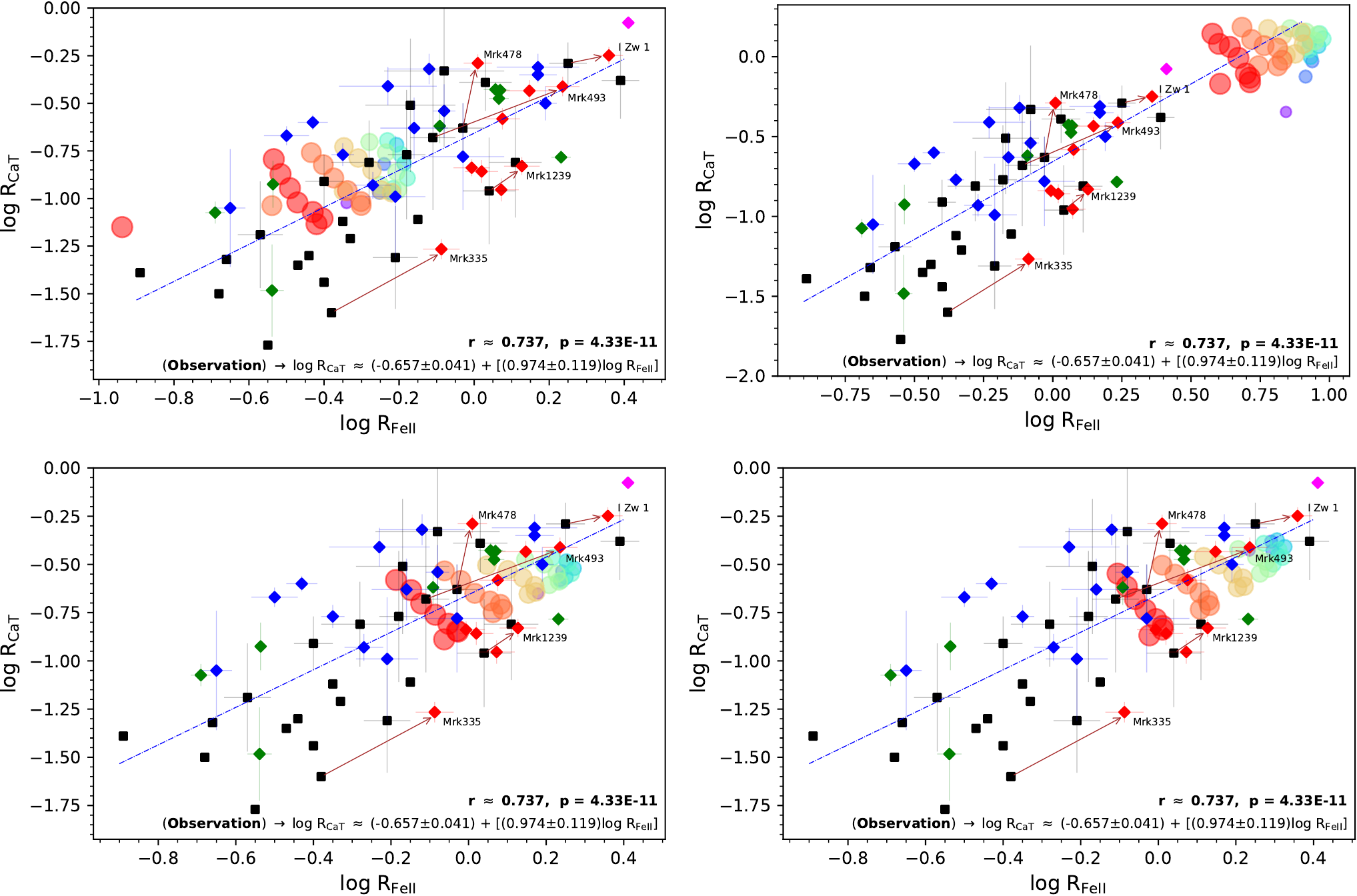}\\
    \vspace{0.1cm}
    \hbox{\hspace{5.5em}\includegraphics[width=0.825\linewidth]{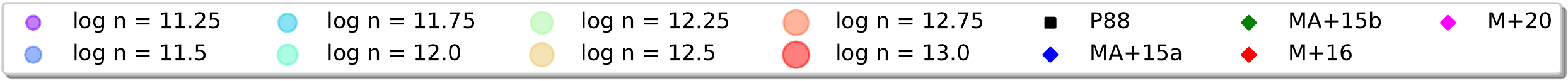}}
    \caption{TOP LEFT: Same as top left panel of Figure \ref{fig:ratioZNHvary}, i.e. at Z = 0.2\zsun{} and N$_{\rm{H}}$ = 10$^{24}\;$cm$^{-2}$. TOP RIGHT: Same as previous panel but at Z = 5\zsun{}. BOTTOM LEFT: Same as bottom left panel of Figure \ref{fig:ratioZNHvary}, i.e. at Z = \zsun{} and N$_{\rm{H}}$ = 10$^{24.5}\;$cm$^{-2}$. BOTTOM RIGHT: Same as previous panel but at N$_{\rm{H}}$ = 10$^{25}\;$cm$^{-2}$. Modelled points from \textmyfont{CLOUDY} are color-coded here as a function of \textit{cloud density} (\n{}).}
    \label{fig:ratioZNHvary_nbased}
\end{figure*}

In Figure \ref{fig:combined}, we combine all the modelled estimates for \rfe{} and \rcat{} for the various cases described until now, i.e., the base model with Z=\zsun{} and $N_H = 10^{24}$ cm$^{-2}$, two additional cases of metallicity at $N_H = 10^{24}$ cm$^{-2}$: (i) Z = 0.2 \zsun{}; and (ii) Z = 5\zsun{}, and, two additional cases of column densities at Z=\zsun{}: (i) $N_H = 10^{24.5}$ cm$^{-2}$; and (ii) $N_H = 10^{25}$ cm$^{-2}$. Combining all the modelled estimates together, the best-fit relation\footnote{The best-fit relation is obtained using the \textit{lm} routine in R reported with standard errors.} between the \rfe{} and \rcat{} is given as:
\begin{equation}
\log \left(\frac{\textmyfont{CaT}}{\textmyfont{H}\beta}\right) \approx (0.78 \pm 0.022) \log \left(\frac{\textmyfont{Fe II}}{\textmyfont{H}\beta}\right) - (0.659 \pm 0.009)
    \label{eq:model}
\end{equation}
which has a Pearson's correlation coefficient, and provide a much better agreement with the observed measurements with a tighter correlation (r$\approx$0.935). \textit{r} $\approx$ 0.935, and a p-value $<$ 2.2$\times 10^{-16}$. The intrinsic scatter for the slope in the modelled best-fit is quite low, highlighting the robustness of the fit. The best-fit slope brings the correlation closer to the slope of the observed best-fit correlation (to 1.63$\sigma$, where $\sigma$=0.119 is the standard deviation for the observed slope). 

\begin{figure*}
    \centering
    \includegraphics[width=0.95\linewidth]{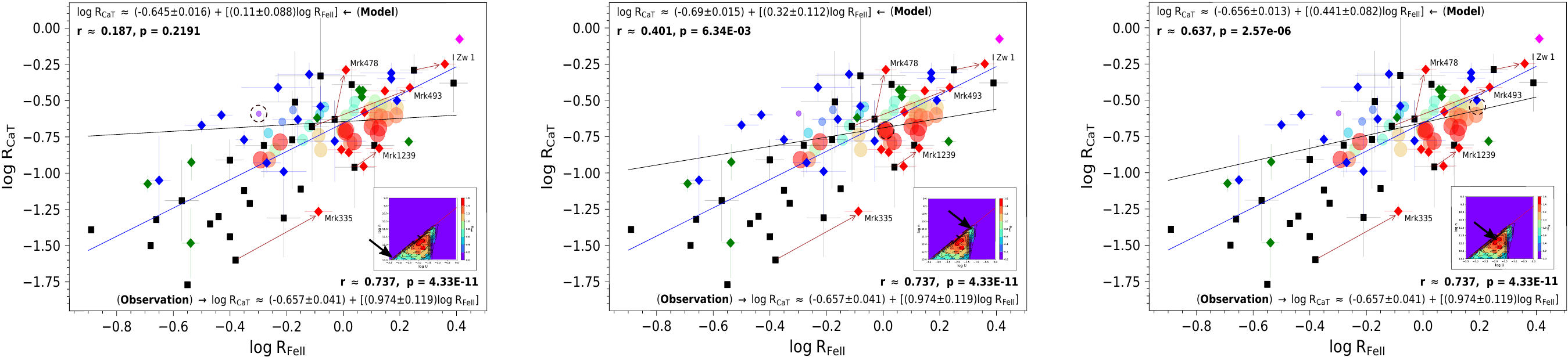}\\
    \vspace{0.1cm}
    \hbox{\hspace{2em}\includegraphics[width=0.95\linewidth]{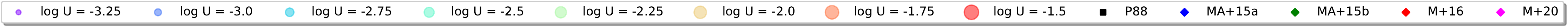}}
    \caption{Effect of un-uniform weightage associated with three individual grid points in the base model (Z=\zsun{}, N$_{\rm{H}}$ = 10$^{24}\;$cm$^{-2}$). Each of the three panels highlight a particular grid point (the location of the grid point is marked in the inset plot and also marked with dashed circle). The grid point in consideration is assigned a weight factor 10, keeping the weights for the remaining points at unity. The best fits for the observational sample (dashed blue) and for the modelled data points (dotted black) and also reported on the plot with corresponding Pearson's correlation coefficients and null probabilities (p-values).}
    \label{fig:testing_uniform_individual}
\end{figure*}

Finally, we note that for all the photoionization models, we assumed equal weights per grid point (see Figure \ref{fig:cat_rfe2}), and hence, an uniform coverage. This assumption affects the best fit obtained from each individual model that is considered in the paper - this is the reason why we opted not to assign the best-fits to the modelled data in the correlation plots where we show single models (Figures \ref{fig:ratio_base_model}, \ref{fig:ratioZNHvary}, \ref{fig:ratioZNHvary_nbased} and \ref{fig:ratioZ1_Un}). For these single models, this is a simplistic assumption, one that is influenced by any \textit{un-uniform} concentration of the ionic species within the BLR cloud. For example, in Figure \ref{fig:testing_uniform_individual}, we show how the best-fit slope gets affected when the concentration of one of the grid point has a preferential weightage (here, the grid point in consideration is assigned a weight factor 10, keeping the weights for the remaining points at unity). In this example, we considered three individual points from the \rfe{}-based panel in our Figure \ref{fig:cat_rfe2} which are also highlighted in the inset plots for each panel of the Figure \ref{fig:testing_uniform_individual}. The effect of this un-uniform weightage can be clearly gauged looking at the change in the slopes for the resulting fits to the modelled data in each case.

\begin{figure*}
    \centering
    \includegraphics[width=0.95\linewidth]{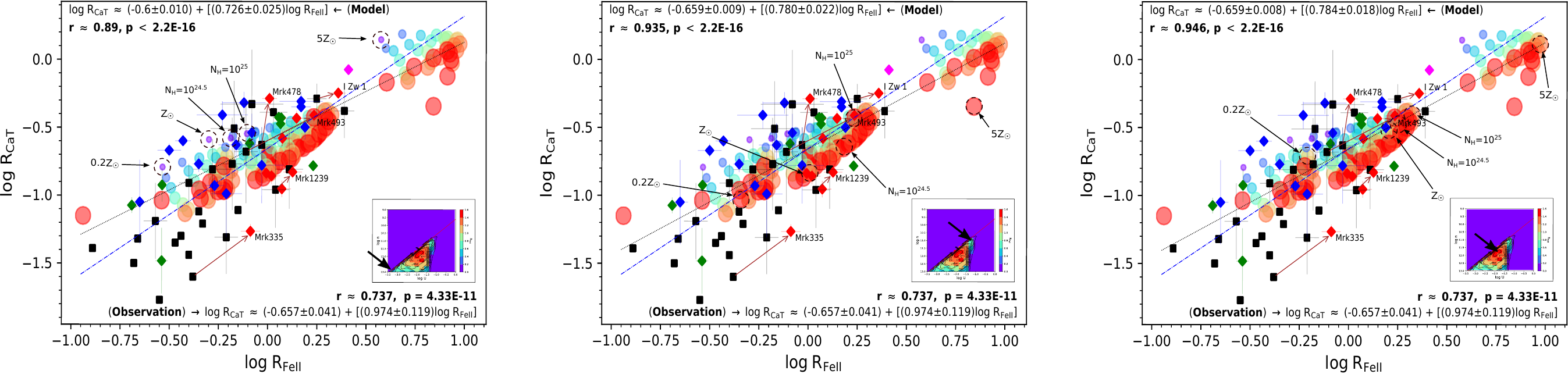}\\
    \vspace{0.1cm}
    \hbox{\hspace{2em}\includegraphics[width=0.95\linewidth]{legend_uniform.pdf}}
    \caption{Effect of un-uniform weightage associated with three individual grid points in all five models combined (the base model with Z=\zsun{} and $N_H = 10^{24}$ cm$^{-2}$, two additional cases of metallicity at $N_H = 10^{24}$ cm$^{-2}$: (i) Z = 0.2 \zsun{}; and (ii) Z = 5\zsun{}, and, two additional cases of column densities at Z=\zsun{}: (i) $N_H = 10^{24.5}$ cm$^{-2}$; and (ii) $N_H = 10^{25}$ cm$^{-2}$). Each of the three panels highlight a particular grid point (same as in Figure \ref{fig:testing_uniform_individual}) for each of the five models (the location of the grid point is marked in the inset plot and also marked with dashed circle). The grid point for each model in consideration are assigned a weight factor 10, keeping the weights for the remaining points at unity. The best fits for the observational sample (dashed blue) and for the combined modelled data points (dotted black) and also reported on the plot with corresponding Pearson's correlation coefficients and null probabilities (p-values).}
    \label{fig:testing_uniform_combined}
\end{figure*}

On the contrary, when all the models are combined (the base model with Z=\zsun{} and $N_H = 10^{24}$ cm$^{-2}$, two additional cases of metallicity at $N_H = 10^{24}$ cm$^{-2}$: (i) Z = 0.2 \zsun{}; and (ii) Z = 5\zsun{}, and, two additional cases of column densities at Z=\zsun{}: (i) $N_H = 10^{24.5}$ cm$^{-2}$; and (ii) $N_H = 10^{25}$ cm$^{-2}$), the effect of these added weights is almost negligible (see modelled best fits in Figure \ref{fig:testing_uniform_combined}), confirming the robustness of the best-fit estimated for the combined models (see Equation \ref{eq:model}).

%%%%%%%%%%%%%%%%%%%%%%%%%%%%%%%%%%%

\section{Discussions} \label{sec:discussions}

We confirm the strong correlation between the strengths of two emission lines species, the optical \feii{} and the NIR \cat{}, both from observations and photoionization modelling. We establish a new best-fit relation for the correlation between the \rcat{} and \rfe{} for an updated catalogue of quasars with measurements for both these species (see Equation \ref{eq:data}). With the inclusion of newer observations, we span a wider and more extended parameter space and we test this with our photoionization models using \textmyfont{CLOUDY} computations parametrized by ionization parameter (\u{}), cloud hydrogen density (\n{}), metallicity (Z) and the size of the cloud constrained by the column density ($\rm{N_H}$); for a representative ionizing continuum shape taken from the broad-band photometric data for \textmyfont{I Zw 1}, a prototypical NLS1 source. These results are qualitative and the nature of the effect of these physical parameters needs to be further investigated on a source-by-source basis.

Recent studies have found connection between high accretion rates and high column densities in the absorbing medium \citep{2017Natur.549..488R}. NLS1s have been shown to have intrinsically higher accretion rates \citep[][and references therein]{2019ApJ...886...42D} and with higher column densities, such as, an additional prospect of the high \feii{} emitting NLS1s being Compton thick sources, i.e. $N_H > 10^{24}$ cm$^{-2}$ going upto $N_H \sim 10^{26}$ cm$^{-2}$ \citep{10.1093/mnras/stv2069}. 

These scenarios need to be tested in the future in the context of the present work. 

\subsection{Effect of a different SED}

\begin{figure}
\centering
\includegraphics[width=0.9\columnwidth]{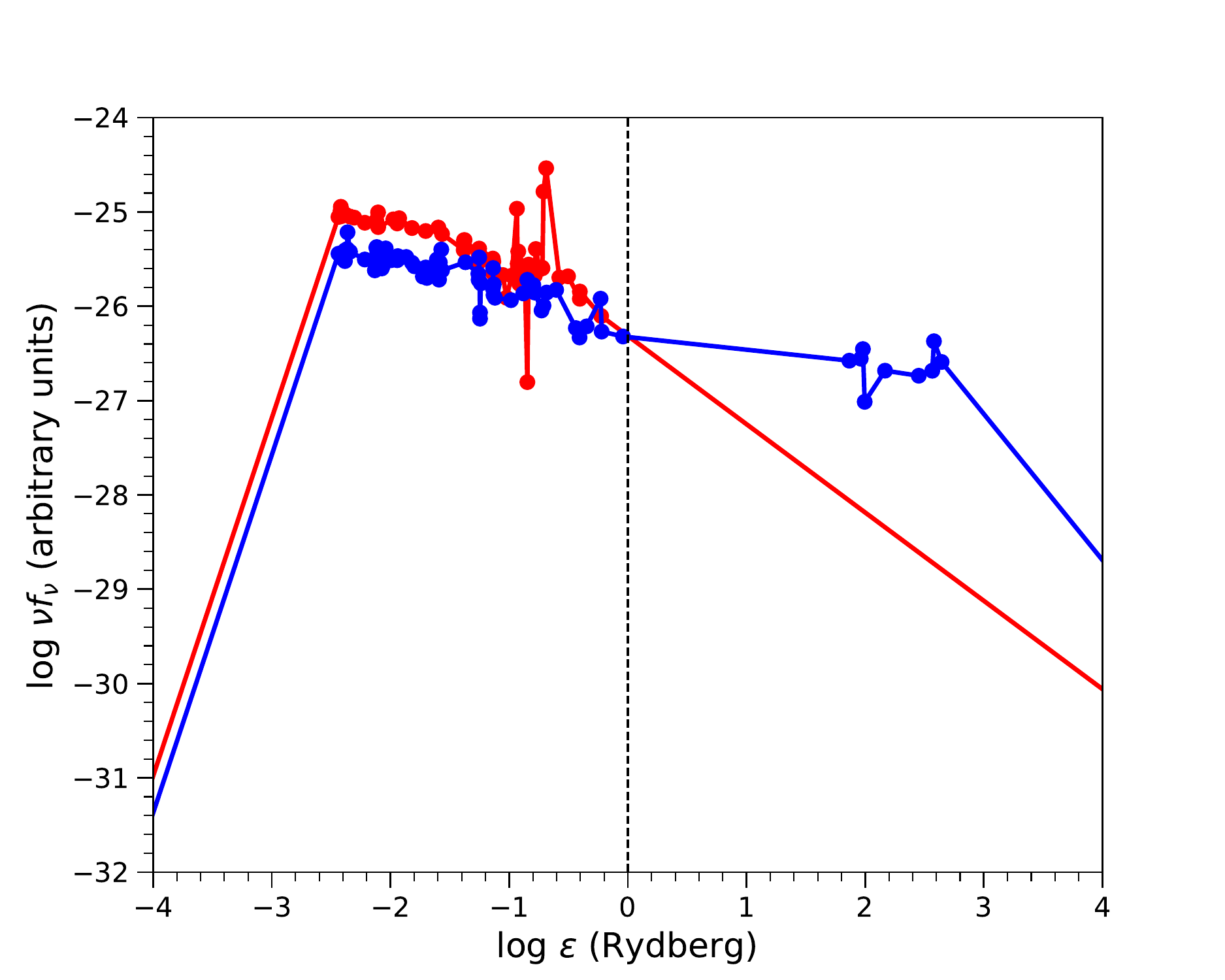}
\caption{Comparison of the spectral energy distributions (SEDs) for \textmyfont{I Zw 1}. The distribution shown in red represents the original SED considered and the other one (in blue) where the SED also includes the X-ray data. The emitted power in arbitrary units is plotted as a function of the photon energy in Ryd. The SEDs have been normalised at $\log\epsilon$ = 0 which is shown with a black dashed line.}
\label{fig:sed_compare}
\end{figure} 

For generality, we used a representative ionizing continuum shape for \textmyfont{I Zw 1}. This is a prototypical NLS1 source which has been extensively studied for decades \citep{ok76,bg92}. Our results confirm the strong correlation between \rcat{} versus \rfe{}, yet there are quite a few outliers from the observational viewpoint, such as Mrk335 and PHL1092 among others. It has been pointed that the dependence of the outer radius of the BLR (i.e. the dust sublimation radius) to the bolometric luminosity \citep{barvainis1987,koshida14}, and, the optical reverberation-mapped BLR radius (equivalent to the inner radius of the BLR) dependence on the monochromatic luminosity \citep{bentz13} is related to the uncertainties on the SED of the source \citep{2020MNRAS.tmp..704N}. We need to account for differences in the SED shapes that has a strong impact on the effective \feii{} production \citep{panda19b,panda19c} and simultaneously on the \cat{} emission.

To test this aspect, we considered the SED for this source including the X-ray emission (shown in blue in Figure \ref{fig:sed_compare}) procured from NED\footnote{\href{http://ned.ipac.caltech.edu/byname?objname=I\%20Zw\%201\&hconst=67.8\&omegam=0.308\&omegav=0.692\&wmap=4\&corr_z=1}{NASA/IPAC Extragalactic Database}.} and compare it against the one used in this paper (obtained from Vizier; shown in red in Figure \ref{fig:sed_compare}). We have incorporated the two SEDs by interpolating the photometric data as they are provided in the databases. We repeated our analysis using the \textit{blue} SED and recovered similar agreement with the observational sample, although the presence of the more energetic photons in excess of 1 Rydberg compared to the \textit{red} SED lifts the overall \rfe{} and \rcat{} estimates (the maximum \rfe{} recovered in the base model with the \textit{blue} SED rises to 1.8 from 1.6, the latter is obtained from the \textit{red} SED. Similarly for \rcat{}, the maximum rises to 0.45 from 0.35). Combining all the modelled estimates together, the best-fit relation between the \rfe{} and \rcat{} for the \textit{blue} SED is given as:
\begin{equation}
\begin{split}
\log \left(\frac{\textmyfont{CaT}}{\textmyfont{H}\beta}\right) \approx & (0.456 \pm 0.036) \log \left(\frac{\textmyfont{Fe II}}{\textmyfont{H}\beta}\right)\\
& - (0.586 \pm 0.010)    
\end{split}
\label{eq:model2}
\end{equation}
which has a Pearson's correlation coefficient, \textit{r} $\approx$ 0.612, and a p-value $<$ 2.2$\times 10^{-16}$. The intrinsic scatter for the modelled best-fit again remains quite low as compared to the observed data although the slope here is much shallower than what was obtained from the models that used the \textit{red} SED (see Figure \ref{fig:blue_combined}). This highlights the importance of the shape of the incident continuum in the study of the BLR, especially for the low-ionization lines' emission. The standard photoionization theory is based on the assumption that the BLR and the distant observer see the same ionizing continuum. This assumption has been questioned in many recent works \citep[][and references therein]{wang14,gaskelletal2019,ferlandetal2020} highlighting the issue of anisotropic radiation from the accretion disk. This necessitates further investigation.

\subsection{Effect of different dust prescriptions}
Keeping rooted to \citet{Nenkova2008} formalism, for a different dust composition, e.g. more carbon grains and/or larger grain sizes, the characteristic $T_{sub}$ could be $\sim$ 2000 K. This gives us a dust sublimation radius: $R_d = 0.189\sqrt{L/10^{45}}$ pc. If we apply this to obtain the parameter space similar to Figure \ref{fig:cat_rfe2}, we will essentially shrink the non-dusty region by a factor $\sim$2.12. This will drop the emission corresponding not only for the lower ionization parameters but also for the low-density models with higher ionisation parameters. In principle, the dusty region consists of dust grains with a distribution of the said properties (sublimation temperature and grain size) which makes the modelling even more complicated. Added to this, is the confirmation of clumpiness in the dusty regions \citep{2010A&A...523A..27H} which contests the relevance of the \citet{Nenkova2008,barvainis1987} relations which assume a smooth distribution of dust in the torus. This needs to be tested in the context of our analyses.

We incorporated the \citet{Nenkova2008} prescription to compute the sublimation radius which inherently assumes a simplified dust grain size, \textit{a} = 0.05 $\mu$m, and, a slight modification in the exponent for the dust temperature, -2.6 instead of -2.8 (the latter is from the original \citet{barvainis1987} assuming a Planckian distribution for the dust grains). Under the assumption of \textit{a} = 0.05 $\mu$m, the ratio of the sublimation radius obtained from both descriptions is only a function of the dust temperature,
\begin{equation}
    \frac{R_{sub,N}}{R_{sub,B}} = 0.973\times \left(\frac{T_{sub}}{1500 K}\right)^{0.2}
\end{equation}
where, the subscripts N = \citet{Nenkova2008} and B = \citet{barvainis1987} prescriptions. The radius obtained from both formalisms becomes equal for a dust temperature around 1780 K, and, only grows by 3\% for a dust temperature around 2000 K.

An increase in the dust grain size can also lead to the shrinking of the dust sublimation radius. \citet{baskin18} have shown that with a high enough particle density ($\sim 10^{10}$ cm$^{-3}$), larger grains of the order of 1$\mu$m can result in a sublimation radius shrunk to the radius of the broad-line region. Such an effect will curtain the emission from the far-side of the BLR cloud. High luminosity plays a vital role here to counter this effect. Contrary to our simplistic approach to determine a singular dust sublimation radius, past works \citep[][and references therein]{baskin18} have pointed towards the dependence of the dust temperature on the gas density, grain size and the chemical composition which implies selective dust grain evaporation. This suggests a dust \textit{sublimation zone} instead of a unique value for the $R_{sub}$. Additionally, for the dust prescription, we have assumed the source luminosity as per \textmyfont{I Zw 1}, i.e. L$_{\rm{bol}} \sim 4.32\times10^{45}$ erg s$^{-1}$. For our observational sample, we have the range of monochromatic luminosity such that, $42.5<\mathrm{log}$ L$_{\rm{5100}}<46.8$, with a mean value at 44.65 which is not far from \textmyfont{I Zw 1}'s luminosity (log $\rm{L_{5100}}$ = 44.54). Thus, scaling the $\rm{L_{5100}}$ for the whole sample using the bolometric correction factor, we get $44.0<\mathrm{log}$ L$_{\rm{bol}}<47.44$. Substituting this range in Equation \ref{eq2} (assuming T$_{sub}$=1500 K), the range of sublimation radii (in parsecs) is \ $0.127\leq R_{sub}\leq 6.654$. The lower limit shrinks the non-dusty BLR by a factor of $\sim$6.5 (with respect to the $R_{sub}$ = 0.83 pc, obtained for \textmyfont{I Zw 1}). This can further curb the overall emission derived from \feii{} and \cat{} considering only the dustless region of the BLR. On the other hand, at the high luminosity end the sublimation radius gets pushed further (by $\sim$8 times), allowing more \feii{} and \cat{} ions to compete for the incoming photons emanated by the accretion disk and hence, increasing the emission from these species. For example, in the case of PHL1092, with $\rm{L_{5100}} \sim 7.638\times 10^{44}\;\rm{erg\;s^{-1}}$ \citep{murilo2020} gives a R$_{sub}$ = 1.138 pc. This is already a 40\% increase in the size of the dustless BLR as compared to what has been assumed in this paper. With the inclusion of a proper SED shape mimicking the PHL1092's emission and this increased size for the dustless BLR, the effective \rfe{} and \rcat{} values may be achievable. We leave this for a future work.

\section{Conclusions} \label{sec:conclusions}
In this article, we confirm the strong correlation between the strengths of two emission lines species, the optical \feii{} and the NIR \cat{}, both from observations and photoionization modelling. With the inclusion of newer observations, we span a wider and extended parameter space and we test this with our photoionization models using \textmyfont{CLOUDY}. We summarize the important conclusions derived from this work as following:
\begin{itemize}
    \item We compile an up-to-date catalogue of 58 sources (including 5 common) with corresponding  \rcat{} and \rfe{} measurements. We then derive the best-fit correlation for \rcat{} versus \rfe{} for our sample (see Equation \ref{eq:data}). With the inclusion of newer data, the best-fit correlation is now consistent with simple proportionality between the two quantities. The best fit slope reported by \citet{martinez-aldamaetal15} was considerably higher (1.33 vs. 1.008 in this paper) but the two values are still consistent within 3$\sigma$ error.
    \item We overlay the modelled data from our photoionization models onto the observed measurements and take this a step further by analysing the optimum parametrization that leads to the maximum emissivity for both these species in terms of (a) ionization parameter; and (b) local cloud density. We find that with our base modelling assumptions, i.e. Z = \zsun{} and column density $N_H = 10^{24}$ cm$^{-2}$, we can quite convincingly explain the mean population amidst the correlation and quantify the range of emission strengths for these two species with this approach.
    \item We further extend our models by testing the effect of metallicity as a way to explain the range of \rcat{} and \rfe{} measurements. We find that the low-\feii{} low-\cat{} emission is well explained by assuming a sub-solar metallicity, such as 0.2\zsun{} keeping the remaining parameters identical to the base model. On the other hand, for the high-end of the correlation, we tested with a super-solar metallicity (Z=5\zsun{}) as shown by previous studies \citep[][and references therein]{panda19c}. We find that the maximum \rfe{} values (e.g. I Zw 1 and PHL1092) can be well explained with fairly low metallicities, i.e. Z $\lesssim$ 5\zsun{}, although above solar values.
    \item Motivated from past studies \citep{perssonferland89}, we also involve the effect of increasing the column density and subsequently find increased \rcat{} and \rfe{} by a factor of $\sim$20\% by enhancing $N_H$ from $10^{24}$ cm$^{-2}$ to $10^{24.5}$ cm$^{-2}$. A further increase of $\sim$20\% is seen by assuming an $N_H = 10^{25}$ cm$^{-2}$ which gives comparable \rcat{} and \rfe{} estimates for the strong-emitting NLS1s. We confirm that an increase in column density relates well with the increased emission for both these species. We clearly find a coupling between the metallicity and column density which leads to increased emission for both \cat{} and \feii{} and bringing the best-fit correlation within 2$\sigma$ of the observed best-fit relation.
    \item Finally, we combine all the modelled estimates for \rfe{} and \rcat{} for the five cases - the base model with Z=\zsun{} and $N_H = 10^{24}$ cm$^{-2}$, two additional cases of metallicity at $N_H = 10^{24}$ cm$^{-2}$: (i) Z = 0.2 \zsun{}; and (ii) Z = 5\zsun{}, and, two additional cases of column densities at Z=\zsun{}: (i) $N_H = 10^{24.5}$ cm$^{-2}$; and (ii) $N_H = 10^{25}$ cm$^{-2}$, and estimate the joint best-fit (see Equation \ref{eq:model}) with a high significance (Pearson's correlation coefficient, r $\approx$ 0.935; and a p-value $< 2.2\times 10^{-16}$). This is in good agreement with the relation given by the observations (see Equation \ref{eq:data}).
\end{itemize}

An increased availability of optical and NIR spectroscopic measurements, especially with the advent of the upcoming ground-based 10-metre-class \citep[e.g. Maunakea Spectroscopic Explorer,][]{2019BAAS...51g.126M} and 40 metre-class \citep[e.g. The European Extremely Large Telescope,][]{2015arXiv150104726E} telescopes; and space-based missions such as the James Webb Space Telescope and the Nancy Grace Roman Space Telescope would further help to accentuate the strong correlation shown by these two ionic species.

\section*{Acknowledgements}
We are grateful to the anonymous referee for his comments and suggestions that helped shape the current version of the manuscript. The project was partially supported by the Polish Funding Agency National Science Centre, project 2017/26/\-A/ST9/\-00756 (MAESTRO  9) and MNiSW grant DIR/WK/2018/12. SP would like to acknowledge the computational facility at CAMK and Dr. Pawe\l{} Cieciel\k{a}g for assistance with the computational cluster. DD acknowledges support from grant PAPIIT UNAM, 113719. We specially thank Alenka Negrete for providing us with the data used in Figure \ref{fig2:fe2_in_blr} in this paper.

\software {\textmyfont{CLOUDY} v17.01 (\citealt{f17}); \textmyfont{MATPLOTLIB}  (\citealt{hunter07}); \textmyfont{NUMPY} (\citealt{numpy}); \textmyfont{R} (\citealt{r_stats})}

%%%%%%%%%%%%%%%%%%%%%%%%%%%%%%%%%%%

\bibliography{main}

\appendix

\section{Product of the cloud density and ionization parameter}
\label{sec:product}

%%%%%%%%%%%%%%%%%%%%%%%%%%%%%%%%%%%
\begin{figure}
\centering
\includegraphics[width=0.8\linewidth]{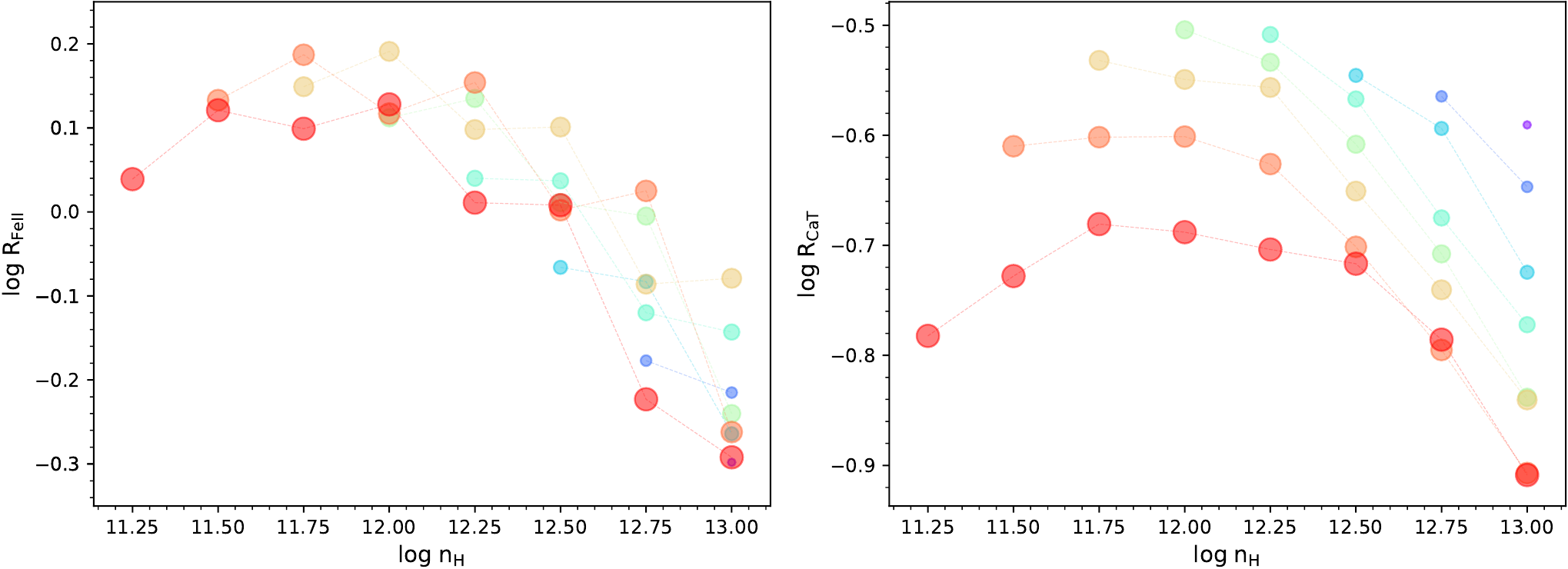}\\
\includegraphics[width=0.45\linewidth]{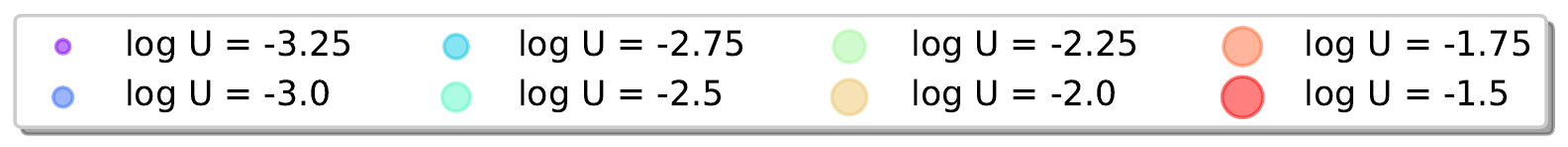}
\caption{Distribution of densities and ionisation parameters for (a) \rfe{}, and (b) \rcat{}. The estimates are for the Base Model as shown in Figure \ref{fig:ratio_base_model}.}
\label{fig:prominence}
\end{figure}

\begin{figure}
    \centering
    \includegraphics[width=0.5\columnwidth]{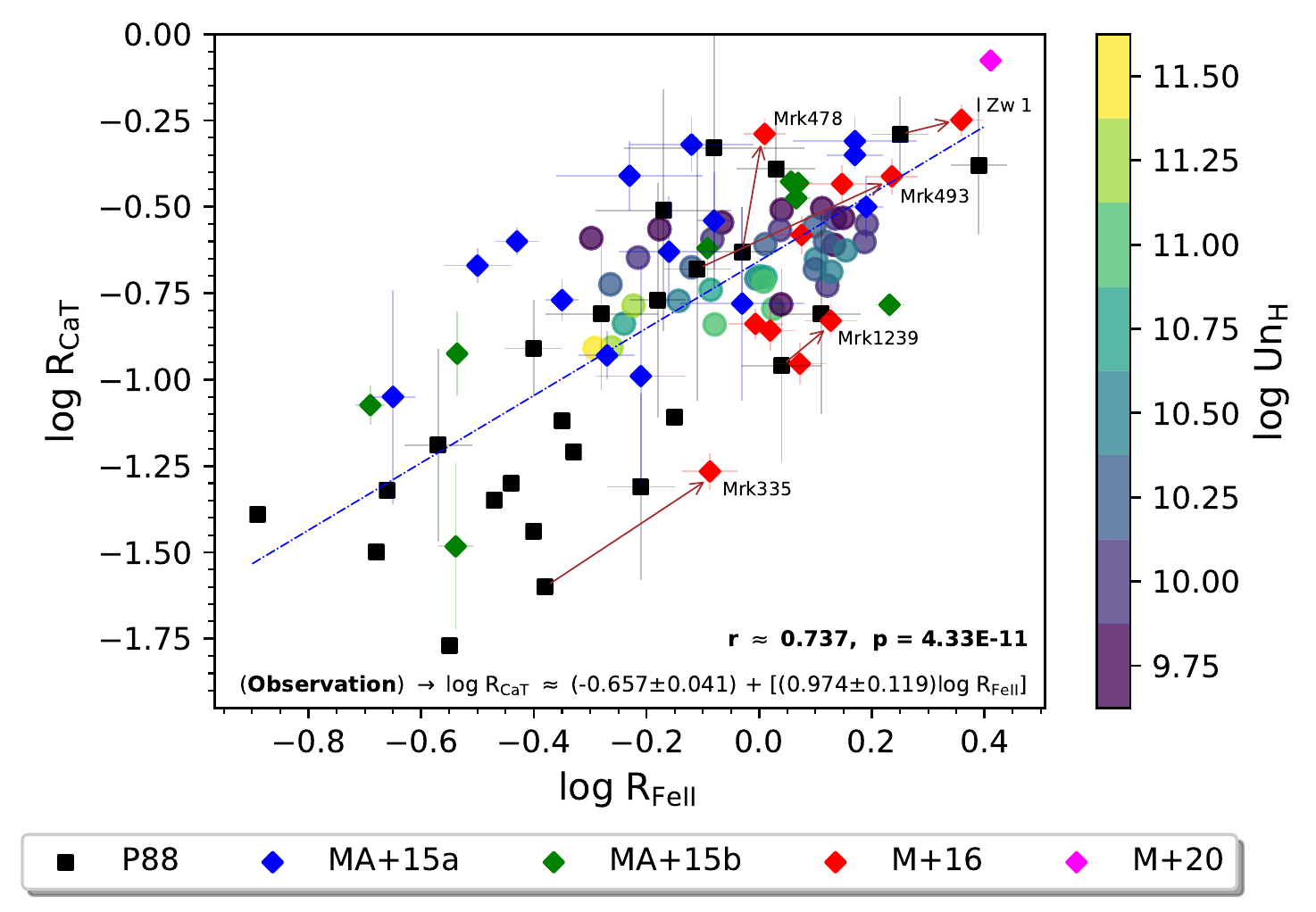}
    \caption{Data points from the photoionization modelling are shown as a function of \u{} $\cdot$ \n{} in log-scale. Rest parameters are identical to the left panel of Figure \ref{fig:ratio_base_model}.}
    \label{fig:ratioZ1_Un}
\end{figure}

In the panels of Figure \ref{fig:ratio_base_model}, we explored the dependence of the effect of the ionization parameter and local cloud density, respectively. The initial idea was to find a way to constrain the sources with respect to these two fundamental physical parameters. However, the dependence on the parameters is clearly non-monotonic which is reflected in the considerable diversity in the distribution when one considers them separately. 

Here, we investigate the coupled distribution between the two parameters. As has been previously explored in \citet{negreteetal12, negrete2014, 2019Atoms...7...18M}, we take the product of the ionization parameter and the local cloud density (\u{}$\cdot$\n{}), i.e. the ionizing photon flux, which can be used as an estimator for the size of the BLR (\rblr{}; see, e.g. \citealt{negreteetal13, 2020arXiv200413113P} and references therein). In Figure \ref{fig:ratioZ1_Un}, we illustrate this with the modelled data from our base model as shown previously (see both panels of Figure \ref{fig:ratio_base_model}) in order to judge if there is a clear trend with respect to \u{}$\cdot$\n{}. Rather interestingly, the highest values for the product of \u{} and \n{} ($\gtrsim$ 11.25 in log-scale) correspond to the lowest values of the \rfe{} ($\log$ \rfe{} $\lesssim$ -0.2). The values for the corresponding \rcat{} are among the lowest as well for these conditions ($\log$ \rcat{} $\lesssim$ -0.75). The highest values for the \rfe{} ($\log$ \rfe{} $\sim$ 0.2) are actually obtained for much smaller values of the product (9.75 $\lesssim \log\,$\u{}\n{} $\lesssim$ 10.25). This is consistent also for the highest values for the \rcat{} ($\log$ \rcat{} $\lesssim$ -0.5) although at slightly lower \u{}$\cdot$\n{}, suggesting larger values for the \rblr{} are required to efficiently produce these ionic species. This implies that, from a strictly photoionization point of view, the two emitting regions are quite closely related in terms of their distance from the central ionizing source, with the \cat{} emitting region further out than the \feii{} region. \citet{panda18b, panda19, panda19b} found that the typical BLR densities to maximise the \feii{} emission strength is achieved with higher densities, typically of the order of 10$^{12}\,\rm{cm^{-3}}$. This, for the maximum \rfe{} value, corresponds to an ionization parameter, $\log$ \u{} $\sim$ -2.0$\pm$0.25 while U $\sim$ -2.25 maximizes the \cat{} emission. This is exactly what we see directly from Figure \ref{fig:cat_rfe2}. There's a clear decrease in the $\log\,$\u{}$\cdot$\n{} along the best-fit relation, i.e., both the \rfe{} and \rcat{} increase with decreasing $\log\,$\u{}$\cdot$\n{}. 

Alternatively, we can look at the dependence on \un{} on maximising the \rfe{} and \rcat{} as shown in the two panels of Figure \ref{fig:prominence} which illustrates the variation in the \rfe{} (left panel) and for \rcat{} (right panel). The color-coding and the increasing point size is identical to Figure \ref{fig:ratio_base_model}. The maximum \rfe{} value (\rfe{} $\approx$ 1.552) is recovered for a BLR density, $\log\,$\n{} = 12 (cm$^{-3}$) and for the ionization parameter, $\log\,$\u{} = -2.0. Note that, these conditions imply a \rcat{} $\approx$ 0.282. Similarly, for the maximum \rcat{} ($\approx$ 0.313), this value of density is identical as above, but for a slightly lower ionization parameter, i.e, $\log\,$\u{} = -2.25, which corresponds to \rfe{} $\approx$ 1.249. The non-monotonic trends are seen for both \rfe{} and \rcat{}. For the given range of cloud densities, we see that the positions of the peak values for \rcat{} per ionization parameter, shift toward lower densities, starting at $\log\,$\n{} = 13.0 (cm$^{-3}$) for a $\log\,$\u{} = -3.25 (although there's only one data point for this case), going till $\log\,$\n{} = 11.75 (cm$^{-3}$) for a $\log\,$\u{} = -1.5. This trend is almost consistent for the \rfe{} and the range of densities for the corresponding peak \rfe{} values are similar to \rcat{}.

Further increase in the density (for $\log\,$\n{} > 12 cm$^{-3}$) leads to a clear decline in both the \rcat{} and \rfe{} values. This is quite straightforward, as we have fixed the maximum column density at $10^{24}$ cm$^{-2}$, the corresponding cloud depth gradually decreases with an increase in cloud density. Hence, the total integrated intensity drops with increasing density. Although, for the densities $\log\,$\n{} $\leq$ 13.0 (cm$^{-3}$), we can observe the peak emission (for both \rcat{} and \rfe{}) when the cloud depth is comparable to the cloud density, i.e. \textit{d} (cm) $\sim$ 10$^{11.75}$ - 10$^{12.25}$ (see Figure \ref{fig:ratioZNHvary_nbased}).

\begin{figure}
    \centering
    \includegraphics[width=0.5\columnwidth]{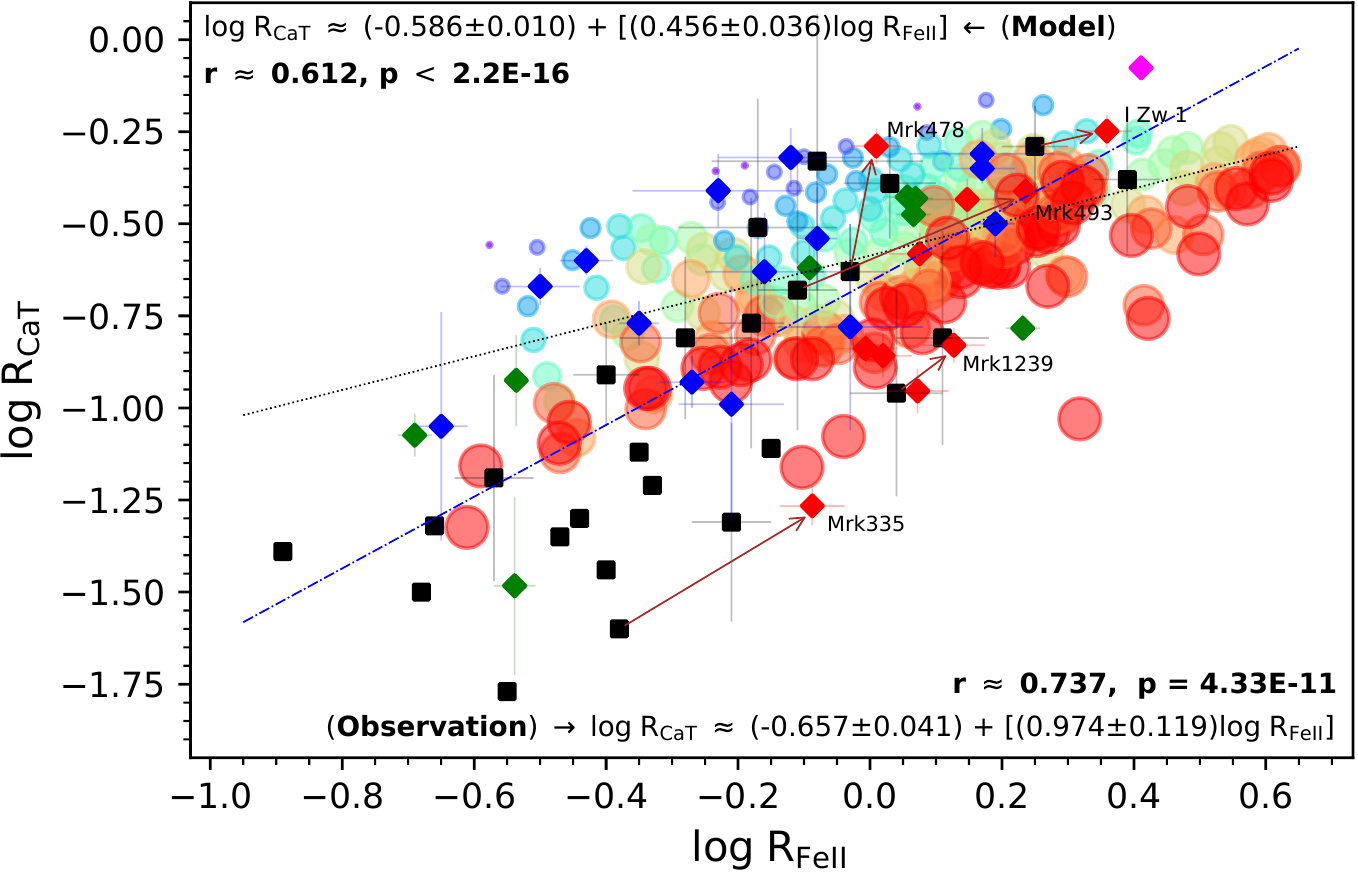}\\
    \vspace{0.1cm}
    \qquad\includegraphics[width=0.5\linewidth]{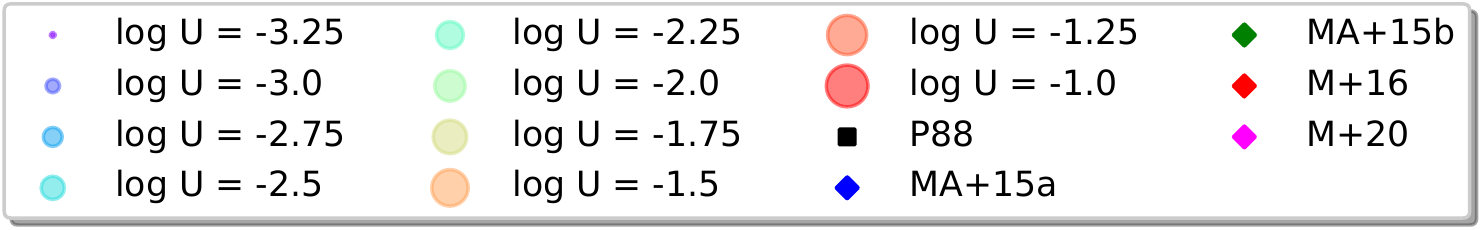}
    \caption{Same as Figure \ref{fig:combined} but the models are generated using the incident SED including the X-ray emission (see the blue continuum in Figure \ref{fig:sed_compare}).}
    \label{fig:blue_combined}
\end{figure}

%%%%%%%%%%%%%%%%%%%Table 1: Physical Parameters
\begin{center}
\scriptsize
\small
\begin{longtable}[c]{cccc}
\caption{Physical parameters for the 58 sources}
\label{tab:table1}\\
 \hline\hline\noalign{\vskip 0.1cm}
Object       & z    & R$_{\rm{FeII}}$     & \rcat{}        \\
\endfirsthead
\endhead

\hline\hline\noalign{\vskip 0.1cm}
\noalign{\vskip 0.1cm} 																			
\multicolumn{4}{c}{\citet{persson1988} Sample} \\ 																													
\hline \noalign{\vskip 0.1cm}

I\,Zw\,1$^{\dag}$	&	0.061	&	1.778$\,\pm\,$0.050	&	0.513$\,\pm\,$0.130	\\
Mrk\,42	&	0.025	&	1.072$\,\pm\,$0.070	&	0.407$\,\pm\,$0.141$^{a}$	\\
Mrk\,478$^{\dag}$	&	0.077	&	0.933$\,\pm\,$0.060	&	0.234$\,\pm\,$0.070	\\
II\,Zw\,136	&	0.063	&	0.661$\,\pm\,$0.050	&	0.170$\,\pm\,$0.133	\\
Mrk\,231	&	0.044	&	2.455$\,\pm\,$0.050	&	0.417$\,\pm\,$0.192	\\
3C\,273	&	0.159	&	0.398$\,\pm\,$0.050	&	0.123$\,\pm\,$0.040	\\
Mrk\,6	&	0.019	&	0.832$\,\pm\,$0.160	&	0.468$\,\pm\,$0.431$^{*}$	\\
Mrk\,486	&	0.039	&	0.269$\,\pm\,$0.060	&	0.065$\,\pm\,$0.042$^{*}$	\\
Mrk\,1239$^{\dag}$	&	0.02	&	1.096$\,\pm\,$0.070	&	0.110$\,\pm\,$0.071$^{a}$	\\
Mrk\,766	&	0.013	&	0.676$\,\pm\,$0.120	&	0.309$\,\pm\,$0.249$^{*}$	\\
Zw\,0033+45	&	0.047	&	0.525$\,\pm\,$0.100	&	0.155$\,\pm\,$0.079$^{*}$	\\
Mrk\,684	&	0.046	&	1.288$\,\pm\,$0.070	&	0.155$\,\pm\,$0.103	\\
Mrk\,335$^{\dag}$	&	0.026	&	0.417$^{u}$	&	0.025$^{u}$	\\
Mrk\,376	&	0.056	&	0.617$\,\pm\,$0.060	&	0.049$\,\pm\,$0.030$^{b}$	\\
Mrk\,493$^{\dag}$	&	0.032	&	0.776$\,\pm\,$0.060	&	0.209$\,\pm\,$0.183$^{*}$	\\
Mrk\,841	&	0.037	&	0.209$^{u}$	&	0.032$^{u}$	\\
Ton\,1542	&	0.063	&	0.363$^{u}$	&	0.05$^{u}$	\\
VII\,Zw\,118	&	0.079	&	0.447$^{u}$	&	0.076$^{u}$	\\
Mrk\,124	&	0.057	&	0.708$^{u}$	&	0.078$^{u}$	\\
Mrk\,9	&	0.04	&	0.398$^{u}$	&	0.036$^{u}$	\\
NGC\,7469	&	0.016	&	0.339$^{u}$	&	0.045$^{*,
u}$	\\
Akn\,120	&	0.034	&	0.468$^{u}$	&	0.062$^{u}$	\\
Mrk\,352	&	0.014	&	0.219$^{u}$	&	0.048$^{*,
u}$	\\
Mrk\,304	&	0.066	&	0.282$^{u}$	&	0.017$^{*,
u}$	\\
Mrk\,509	&	0.034	&	0.129$^{u}$	&	0.041$^{u}$	\\

\hline\noalign{\vskip	0.1cm}						
\noalign{\vskip	0.1cm}

\multicolumn{4}{c}{\citet{martinez-aldamaetal15}	Sample}	\\					
\hline	\noalign{\vskip	0.1cm}					
							
HE0005-2355	&	1.412	&	0.933$\,\pm\,$0.237	&	0.166$\,\pm\,$0.107	\\
HE0035-2853	&	1.638	&	1.479$\,\pm\,$0.170	&	0.447$\,\pm\,$0.041	\\
HE0043-2300	&	1.54	&	0.316$\,\pm\,$0.044	&	0.214$\,\pm\,$0.025	\\
HE0048-2804	&	0.847	&	0.617$\,\pm\,$0.114	&	0.102$\,\pm\,$0.075	\\
HE0058-3231	&	1.582	&	0.589$\,\pm\,$0.176	&	0.389$\,\pm\,$0.090	\\
HE0203-4627	&	1.438	&	0.759$\,\pm\,$0.192	&	0.479$\,\pm\,$0.088	\\
HE0248-3628	&	1.536	&	0.372$\,\pm\,$0.034	&	0.251$\,\pm\,$0.023	\\
HE1349+0007	&	1.444	&	0.692$\,\pm\,$0.143	&	0.234$\,\pm\,$0.086	\\
HE1409+0101	&	1.65	&	1.549$\,\pm\,$0.107	&	0.316$\,\pm\,$0.066	\\
HE2147-3212	&	1.543	&	1.479$\,\pm\,$0.375	&	0.490$\,\pm\,$0.079	\\
HE2202-2557	&	1.535	&	0.537$\,\pm\,$0.062	&	0.117$\,\pm\,$0.019	\\
HE2340-4443	&	0.922	&	0.224$\,\pm\,$0.021	&	0.089$\,\pm\,$0.064	\\
HE2349-3800	&	1.604	&	0.832$\,\pm\,$0.057	&	0.288$\,\pm\,$0.093	\\
HE2352-4010	&	1.58	&	0.447$\,\pm\,$0.031	&	0.170$\,\pm\,$0.023	\\
							
\hline\noalign{\vskip	0.1cm}						
\noalign{\vskip	0.1cm}						
							
\multicolumn{4}{c}{\citet{martinez-aldamaetal15b}	Sample}	\\					
\hline	\noalign{\vskip	0.1cm}					
							
HE0349-5249	&	1.541	&	1.704$\,\pm\,$0.102	&	0.165$\,\pm\,$0.014	\\
HE0359-3959	&	1.521	&	1.173$\,\pm\,$0.070	&	0.371$\,\pm\,$0.031	\\
HE0436-3709	&	1.445	&	1.164$\,\pm\,$0.070	&	0.335$\,\pm\,$0.028	\\
HE0507-3236	&	1.577	&	0.291$\,\pm\,$0.006	&	0.119$\,\pm\,$0.034	\\
HE0512-3329	&	1.587	&	0.810$\,\pm\,$0.017	&	0.240$\,\pm\,$0.050	\\
HE0926-0201	&	1.682	&	1.139$\,\pm\,$0.082	&	0.374$\,\pm\,$0.033	\\
HE1039-0724	&	1.458	&	0.289$\,\pm\,$0.021	&	0.033$\,\pm\,$0.018	\\
HE1120+0154	&	1.472	&	0.204$\,\pm\,$0.012	&	0.084$\,\pm\,$0.011	\\
							
\hline\noalign{\vskip	0.1cm}						
\noalign{\vskip	0.1cm}						
\multicolumn{4}{c}{\citet{murilo2016}	Sample}	\\					
\hline	\noalign{\vskip	0.1cm}					
							
1H\,1934-063	&	0.011	&	1.404$\,\pm\,$0.223	&	0.368$\,\pm\,$0.047	\\
1H\,2107-097	&	0.027	&	1.047$\,\pm\,$0.106	&	0.139$\,\pm\,$0.019	\\
I\,Zw\,1$^{\dag}$	&	0.061	&	2.286$\,\pm\,$0.199	&	0.564$\,\pm\,$0.058	\\
Mrk\,1044	&	0.016	&	1.181$\,\pm\,$0.127	&	0.111$\,\pm\,$0.016	\\
Mrk\,1239$^{\dag}$	&	0.019	&	1.340$\,\pm\,$0.147	&	0.148$\,\pm\,$0.016	\\
Mrk\,335$^{\dag}$	&	0.026	&	0.818$\,\pm\,$0.092	&	0.054$\,\pm\,$0.007	\\
Mrk\,478$^{\dag}$	&	0.078	&	1.023$\,\pm\,$0.089	&	0.514$\,\pm\,$0.056	\\
Mrk\,493$^{\dag}$	&	0.032	&	1.721$\,\pm\,$0.179	&	0.387$\,\pm\,$0.046	\\
PG\,1448+273	&	0.065	&	1.189$\,\pm\,$0.129	&	0.262$\,\pm\,$0.034 \\
Tons\,180	&	0.062	&	0.985$\,\pm\,$0.110	&	0.145$\,\pm\,$0.015	\\
							
\hline\noalign{\vskip	0.1cm}						
\noalign{\vskip	0.1cm}

\multicolumn{4}{c}{\citet{murilo2020}}	\\						
\hline	\noalign{\vskip	0.1cm}					
							
PHL\,1092	&	0.394	&	2.576$\,\pm\,$0.108	&	0.839$\,\pm\,$0.038	

\end{longtable}
\end{center}
\footnotesize{{\sc Notes.} Columns are as follows: (1) Object name. (2) Redshift. (3) \rfe{} = EW(FeII$_{4434-4684}$)/EW(H$\beta$). (4) \cat{} = EW(\cat{})/EW(H$\beta$). In column 1 ($\dag$) symbol represents the 5 common sources in \cite{persson1988} and \cite{murilo2016} samples. In column 4,
 $*$ symbol indicates sources with a possible host galaxy contribution in the \cat\ emission lines, $a$ letter indicates the presence of a central dip in the CaII emission lines, $b$ letter marks a questionable \cat\ detection, and $u$ letter indicates an upper limit in \hb\ and \cat\ intensity. } 

\end{document}